\documentclass[11pt]{article}
\usepackage{epsfig}
\newcommand{\sect}[1]{ \section{#1} \setcounter{equation}{0} }

\newcommand{\Dslash}{D \! \! \! \! /} 
 
\newcommand{\pslash}{p \! \! \! /} 
\newcommand{\qslash}{q \! \! \! /}

\newcommand{\third}{\mbox{\small{$\frac{1}{3}$}}}

\newcommand{\MSbar}{\overline{\mbox{MS}}} 
\newcommand{\MSbars}{\overline{\mbox{\footnotesize{MS}}}} 
\newcommand{\MOMggg}{\mbox{MOMggg}}
\newcommand{\MOMgggs}{\mbox{\footnotesize{MOMggg}}}
\newcommand{\MOMh}{\mbox{MOMh}}
\newcommand{\MOMhs}{\mbox{\footnotesize{MOMh}}}
\newcommand{\MOMq}{\mbox{MOMq}}
\newcommand{\MOMqs}{\mbox{\footnotesize{MOMq}}}
\newcommand{\MOMi}{\mbox{MOMi}}
\newcommand{\MOMis}{\mbox{\footnotesize{MOMi}}}
\newcommand{\Nc}{N_{\!c}}
\newcommand{\Nf}{N_{\!f}}
\newcommand{\NF}{N_{\!F}}
\newcommand{\NA}{N_{\!A}}
\newcommand{\Nda}{N^d_{\!A}}
\newcommand{\Noda}{N^o_{\!A}}

\begin{document}
\title{MOM renormalization group functions in the maximal abelian gauge}
\author{J.M. Bell \& J.A. Gracey, \\ Theoretical Physics Division, \\ 
Department of Mathematical Sciences, \\ University of Liverpool, \\ P.O. Box 
147, \\ Liverpool, \\ L69 3BX, \\ United Kingdom.} 
\date{} 
\maketitle 

\vspace{5cm} 
\noindent 
{\bf Abstract.} The one loop $3$-point vertex functions of QCD in the maximal 
abelian gauge (MAG) are evaluated at the fully symmetric point at one loop. As 
a consequence the theory is renormalized in the various momentum (MOM) schemes
which are defined by the trivalent vertices, as well as in the $\MSbar$ scheme.
From these the {\em two} loop renormalization group functions in the MOM 
schemes are derived using the one loop conversion functions. In parallel we 
repeat the analysis for the Curci-Ferrari gauge which corresponds to the MAG in
a specific limit. The relation between the $\Lambda$ parameters in different 
schemes is also provided. 

\vspace{-17cm}
\hspace{13.5cm}
{\bf LTH 985}

\newpage

\sect{Introduction.} 

One of the outstanding problems in quantum field theory is to understand the
mechanism behind quark and gluon confinement. The former are the building
blocks of hadronic states while the latter are the quanta which mediate the
strong nuclear force. Unlike other fundamental particles in the standard model
neither quarks nor gluons are seen in nature as isolated states. Though at high
energy quarks behave as asymptotically free entities and to all intents and 
purposes are seen through their interaction within deep inelastic scattering, 
for example. However, this high energy property of asymptotic fundamentality 
does not persist at low energies. Instead infrared slavery dominates and single
free quarks cannot be isolated. In other words the full quark and gluon 
propagators do not have simple poles at a zero or non-zero mass. There have 
been many attempts to explain the absence of free quark and gluon states. For 
background see, for instance, the review article \cite{1}. One framework which 
has received attention is that where the infrared dynamics is based on an
abelian theory involving magnetic monopoles, \cite{1,2,3,4,5}. In a parallel of
what occurs in superconductivity, colour charge is confined when the monopoles
condense to produce an Abrikosov-Nielsen-Oleson string, \cite{1,2,3,4}. A main 
key in the whole picture is the underlying abelian structure within a theory 
which has a non-abelian colour group. Thus the actual mathematical structure of
the colour group of the underlying quantum field theory describing the strong 
force plays an important role, \cite{5}. This is either Yang-Mills theory which
describes purely gluons or Quantum Chromodynamics (QCD) when quarks are 
included and involve the non-abelian Lie group $SU(3)$. The abelian monopoles 
are associated with the quanta derived from the centre of the colour group. 
These are believed to dominate the infrared dynamics, \cite{2,3,4,5}. In other 
words the contribution from the remaining off-diagonal sector quanta are 
negligible. 

To understand this picture further from a quantum field theory viewpoint 
requires accessing each sector of the colour group. However, in the canonical
formulation of Yang-Mills or QCD using a linear covariant gauge fixing, one has
no direct access to examining separate centre or off-diagonal gluon dynamics. 
Moreover, one requires techniques to study the field theory non-perturbatively.
One useful method is that of Schwinger-Dyson equations. In this approach the 
aim is to solve the tower of coupled $n$-point functions, usually in a 
particular approximation, that allows clean access to the problem at hand. 
Though for an abelian monopole analysis one has to have a way of making contact
with the centre directly. One way of achieving this is to choose an appropriate
gauge fixing. One such gauge is the maximal abelian gauge (MAG), \cite{4,6,7}. 
It is a nonlinear covariant gauge fixing where the centre and off-diagonal 
gluons are gauge fixed differently, \cite{4,6,7}. While this has been used in 
Schwinger-Dyson analyses, such as \cite{8,9,10}, and several lattice studies, 
such as \cite{11,12,13,14,15,16}, there has not been as much activity in the
MAG compared with the Landau gauge. Encouraging results have emerged 
such as differing infrared behaviours of the centre and off-diagonal gluon and 
ghost propagators. Although the focus has primarily been on $2$-point 
functions, more recently Landau gauge studies have turned to vertex functions 
and specifically $3$-point functions, \cite{17}. These functions have been 
studied at several momentum configurations. The two main ones are the 
asymmetric and the symmetric points. The former is easier to simulate on the 
lattice whereas the latter has relatively noisier signals. However, at the 
symmetric point there are no infrared problems since the momentum configuration
is non-exceptional in contrast to the asymmetric point.

The issue of the subtraction point of the vertices is related to the area of 
renormalization schemes. In \cite{18} the momentum (MOM) subtraction schemes
were introduced for the $3$-point vertices of QCD where the focus was on
linear covariant gauges. This family of schemes are mass dependent 
renormalization schemes which are physical. The actual subtraction is such that
after renormalization the Lorentz channel of the $2$ and $3$-point functions 
containing the divergences is unity at the renormalization point. This original
analysis of \cite{18} was extended to the next loop order recently in 
\cite{19}. Given that the lattice measures vertex functions non-perturbatively 
and requires matching to the high energy behaviour, the more loop order 
information available at high energy allows one to have reduced error estimates
on infrared measurements. In addition Schwinger-Dyson analyses of Green's 
functions requires matching. This was in part the motivation of \cite{19}. 
Based on this and the interest in the infrared structure of QCD in the MAG, it 
is therefore the purpose of this article to provide an analysis of the 
$3$-point vertex functions of QCD in the MAG at one loop. This will extend 
earlier work on the MAG for various colour groups, 
\cite{20,21,22,23,24,25,26,27,28,29}. Moreover, it will be a complete parallel 
of \cite{18} for the linear covariant gauge fixings. We will also provide the 
symmetric point $3$-point vertices both in the $\MSbar$ scheme as well as the 
various MOM schemes associated with the trivalent vertices. One major 
consequence is that the two loop renormalization group functions will be 
deduced in the MOM schemes. This is because the mappings of the various 
parameters between the schemes are derived from the one loop analysis. Hence 
these conversion functions between schemes are used in conjuction with the 
renormalization group equation and the known two loop $\MSbar$ renormalization 
group functions, \cite{30}. As checks on the results we will compare with the 
nonlinear covariant gauge known as the Curci-Ferrari gauge, \cite{31}. In a 
certain limit the MAG is equivalent to this gauge and we have performed the 
full analysis in the Curci-Ferrari gauge. By taking the limit from the MAG we 
will be able to verify agreement. Indeed the Curci-Ferrari gauge is of interest
in its own right as it has a special feature. Originally it was noted in 
\cite{31} that a BRST invariant gluon mass could be included in the Lagrangian.
Clearly it is not gauge invariant but it was regarded as a useful tool for 
potentially modelling gluon mass. Indeed lattice and Schwinger-Dyson analyses 
in recent years have indicated that the Landau gauge gluon propagator freezes 
in the infrared to a finite non-zero value. The initial observations in this 
respect can be found in \cite{32,33,34,35,36,37,38,39,40}. This freezing would 
correspond to some type of effective gluon mass. Moreover, studies in the MAG 
on the lattice suggest a similar phenomena but with differing masses for centre
and off-diagonal gluons, \cite{13,14,15,16}. This splitting of masses in the 
infrared is believed to be symptomatic of the dominance of the abelian sector. 

The paper is organized as follows. We provide all the relevant background to
the MAG in section $2$ including group theory identities we use when the 
centre is identified, the renormalization group scheme conversion functions and
the computational setup for the symmetric point analysis. The subsequent
sections are devoted to the explicit results. The $\MSbar$ amplitudes are given
in section $3$. The two loop renormalization group functions, amplitudes and 
conversion functions for the three MOM schemes are given respectively in
sections $4$, $5$ and $6$. The results for the related Curci-Ferrari gauge are
presented in section $7$ with the derivation of the relation between the
$\Lambda$ parameters given in section $8$. Concluding remarks are given in 
section $9$. An appendix collects the tensor basis of the vertex functions and 
the explicit forms of the associated projection matrices. 

\sect{Background.}

We devote this section to reviewing the key properties of the MAG as well as 
the calculational techniques which we use. First, as noted the MAG is a
particular gauge fixing where the gluons are allocated to two parts of the
colour group, \cite{4,6,7}. Those deriving from the centre are named diagonal 
or centre gluons while those which are not part of this abelian subgroup are 
termed off-diagonal. Given this we use the same notation as \cite{29,30} using
letters $a$, $b$ and $c$ as off-diagonal indices but $i$, $j$, $k$ and $l$ as 
indices for gluons and other fields associated with the centre. Capital letters
are reserved for the adjoint indices of the full colour group. Thus we 
decompose the group valued gauge field, ${\cal A}_\mu$~$=$~$A^A_\mu T^A$ into
\begin{equation}
{\cal A}_\mu ~=~ A^a_\mu T^a ~+~ A^i_\mu T^i
\end{equation}
where $T^A$ are the group generators. As we will be summing over colour indices
we define the dimensions of the diagional sector as $\Nda$ in the adjoint
representation and $\Noda$ for the off-diagonal sector. Thus 
$1$~$\leq$~$i$~$\leq$~$\Nda$, $1$~$\leq$~$a$~$\leq$~$\Noda$ and 
$1$~$\leq$~$A$~$\leq$~$\NA$ where $\NA$ is the dimension of the adjoint
representation in the full group. The dimension of the fundamental is $\NF$. So
\begin{equation}
\Nda ~+~ \Noda ~=~ \NA ~.
\end{equation}
As an example for $SU(\Nc)$ we have $\Nda$~$=$~$\Nc-1$ and
$\Noda$~$=$~$\Nc(\Nc-1)$. Though we will work throughout with an arbitrary
colour group and only specify $SU(3)$ in certain cases. In this notation the
gauge invariant QCD Lagrangian is, \cite{29}, 
\begin{equation}
L ~=~ -~ \frac{1}{4} G^A_{\mu\nu} G^{A \, \mu\nu} ~+~ i \bar{\psi} \Dslash
\psi ~+~ L_{\mbox{\footnotesize{gf}}}
\end{equation}
where $G^A_{\mu\nu}$ is the usual field strength and there are $\Nf$ massless
quarks $\psi$. Translating this to the MAG situation the field strength splits 
into two terms since 
\begin{equation}
L ~=~ -~ \frac{1}{4} G^a_{\mu\nu} G^{a \, \mu\nu} ~-~
\frac{1}{4} G^i_{\mu\nu} G^{i \, \mu\nu} ~+~ i \bar{\psi} \Dslash \psi ~+~
L^{\mbox{\footnotesize{MAG}}}_{\mbox{\footnotesize{gf}}} ~.
\end{equation}
The main difference between this Lagrangian and the canonical covariant gauge
fixing term is that in 
$L^{\mbox{\footnotesize{MAG}}}_{\mbox{\footnotesize{gf}}}$ the prescription to
fix the gauge for the off-diagonal gluons is different from that of the 
diagonal gluons, \cite{4,6,7}. As this construction has been discussed 
elsewhere we record the full gauge fixed MAG as
\begin{eqnarray}
L^{\mbox{\footnotesize{MAG}}}_{\mbox{\footnotesize{gf}}} &=& 
-~ \frac{1}{2\alpha} \left( \partial^\mu A^a_\mu \right)^2 
- \frac{1}{2\bar{\alpha}} \left( \partial^\mu A^i_\mu \right)^2 
+ \bar{c}^a \partial^\mu \partial_\mu c^a
+ \bar{c}^i \partial^\mu \partial_\mu c^i \nonumber \\
&& +~ g \left[ f^{abk} A^a_\mu \bar{c}^k \partial^\mu c^b
- f^{abc} A^a_\mu \bar{c}^b \partial^\mu c^c 
- \frac{1}{\alpha} f^{abk} \partial^\mu A^a_\mu A^b_\nu A^{k \, \nu}
- f^{abk} \partial^\mu A^a_\mu c^b \bar{c}^k
\right. \nonumber \\
&& \left. ~~~~~~
- \frac{1}{2} f^{abc} \partial^\mu A^a_\mu \bar{c}^b c^c 
- 2 f^{abk} A^k_\mu \bar{c}^a \partial^\mu \bar{c}^b
- f^{abk} \partial^\mu A^k_\mu \bar{c}^b c^c \right] \nonumber \\
&& +~ g^2 \left[ f_d^{acbd} A^a_\mu A^{b \, \mu} \bar{c}^c c^d
- \frac{1}{2\alpha} f_o^{akbl} A^a_\mu A^{b \, \mu} A^k_\nu A^{l \, \nu}
+ f_o^{adcj} A^a_\mu A^{j \, \mu} \bar{c}^c c^d \right.
\nonumber \\
&& \left. ~~~~~~~
- \frac{1}{2} f_o^{ajcd} A^a_\mu A^{j \, \mu} \bar{c}^c c^d
+ f_o^{ajcl} A^a_\mu A^{j \, \mu} \bar{c}^c c^l
+ f_o^{alcj} A^a_\mu A^{j \, \mu} \bar{c}^c c^l \right.
\nonumber \\
&& \left. ~~~~~~~
- f_o^{cjdi} A^i_\mu A^{j \, \mu} \bar{c}^c c^d
- \frac{\alpha}{4} f_d^{abcd} \bar{c}^a \bar{c}^b c^c c^d
- \frac{\alpha}{8} f_o^{abcd} \bar{c}^a \bar{c}^b c^c c^d
+ \frac{\alpha}{8} f_o^{acbd} \bar{c}^a \bar{c}^b c^c c^d \right. \nonumber \\
&& \left. ~~~~~~~
- \frac{\alpha}{4} f_o^{abcl} \bar{c}^a \bar{c}^b c^c c^l
+ \frac{\alpha}{4} f_o^{acbl} \bar{c}^a \bar{c}^b c^c c^l
- \frac{\alpha}{4} f_o^{albc} \bar{c}^a \bar{c}^b c^c c^l
+ \frac{\alpha}{2} f_o^{akbl} \bar{c}^a \bar{c}^b c^k c^l \right] \,.
\label{lagmag}
\end{eqnarray}
It is worth noting at this stage we are basing this on the more general
modified MAG discussed in \cite{21}. Though the interpolating parameter,
$\zeta$, which is apparent in \cite{21} and \cite{22} is set to the specific
value for the MAG itself. Its interpolating property is not relevant for this
article. In addition to this, given the nature of this construction we need to 
make comments relevant to our analysis. First, there are two gauge parameters, 
$\alpha$ and $\bar{\alpha}$. The latter is the parameter associated with the 
centre gluons and as such only appears in the quadratic term. It is necessary 
in order to construct the centre gluon propagator and is set to zero 
thereafter. In other words that sector is gauge fixed in the Landau gauge,
\cite{22,29}. For the off-diagonal gluons the gauge parameter appears in the 
interactions as well as the quadratic term. Moreover, it cannot be set to zero 
after the propagator has been constructed since several interactions would have
singular couplings. Thus a non-zero $\alpha$ is retained throughout. Though we 
note that aside from the gauge parameter renormalization all the other 
renormalization group functions are finite in the $\alpha$~$\rightarrow$~$0$ 
limit. For the quartic terms we use a compact notation for the structure 
functions, \cite{30}, 
\begin{equation}
f_d^{ABCD} ~=~ f^{iAB} f^{iCD} ~~~,~~~
f_o^{ABCD} ~=~ f^{eAB} f^{eCD} ~.
\end{equation}
In other words the subscript denotes whether the summed index is from the
centre or off-diagonal sector. In this respect it is worth noting one 
consequence of the Lie algebra. If 
\begin{equation}
\left[ T^A , T^B \right] ~=~ i f^{ABC} T^C 
\end{equation}
then
\begin{equation}
f^{ijk} ~=~ 0 ~~~,~~~ f^{ijc} ~=~ 0 
\end{equation}
and
\begin{equation}
\left[ T^a , T^j \right] ~=~ i f^{ajc} T^c ~. 
\end{equation}
These are important when it comes to performing the group theory associated 
with Feynman diagrams. In addition to the $\alpha$ dependence in (\ref{lagmag})
the gauge fixing requires Faddeev-Popov ghosts, $c^A$. These are associated 
with each colour sector. It is worth noting that while ordinarily an abelian 
gauge theory does not have coupled ghosts this statement only applies to the 
case where the gauge fixing is linear. For instance, in the 't~Hooft-Veltman 
gauge, \cite{41}, there are interacting Faddeev-Popov ghosts. The situation is 
the same here in that the MAG, being a nonlinear gauge fixing, produces centre 
ghosts which couple in a non-trivial fashion. Moreover, there are quartic ghost
terms. These together with the other interactions do not spoil 
renormalizability which has been established in \cite{20,22,23,24,26,29}. As 
part of the renormalization we note that the renormalization constants for the 
fields and the parameters are  
\begin{eqnarray}
A^{a \, \mu}_{\mbox{\footnotesize{o}}} &=& \sqrt{Z_A} \, A^{a \, \mu} ~,~
A^{i \, \mu}_{\mbox{\footnotesize{o}}} ~=~ \sqrt{Z_{A^i}} \, A^{i \, \mu} ~,~
c^a_{\mbox{\footnotesize{o}}} ~=~ \sqrt{Z_c} \, c^a ~,~
\bar{c}^a_{\mbox{\footnotesize{o}}} ~=~ \sqrt{Z_c} \, \bar{c}^a \nonumber \\
c^i_{\mbox{\footnotesize{o}}} &=& \sqrt{Z_{c^i}} \, c^i ~~,~~
\bar{c}^i_{\mbox{\footnotesize{o}}} ~=~ \frac{\bar{c}^i}{\sqrt{Z_{c^i}}} ~~,~~
\psi_{\mbox{\footnotesize{o}}} ~=~ \sqrt{Z_\psi} \psi \nonumber \\
g_{\mbox{\footnotesize{o}}} &=& \mu^\epsilon Z_g \, g  ~~,~~
\alpha_{\mbox{\footnotesize{o}}} ~=~ Z^{-1}_\alpha Z_A \, \alpha  ~~,~~
\bar{\alpha}_{\mbox{\footnotesize{o}}} ~=~ Z^{-1}_{\alpha^i} Z_{A^i} \,
\bar{\alpha} 
\label{zdef}
\end{eqnarray}
where the index $i$ on objects in the subscript are to indicate the centre
sector and there is no summation over this label when it is repeated. Bare
quantities are denoted by the subscript ${}_{\mbox{\footnotesize{o}}}$. We use
the same conventions as \cite{30}. In particular we dimensionally regularize in
$d$~$=$~$4$~$-$~$2\epsilon$ dimensions where $\epsilon$ is the regularizing
parameter and the mass scale $\mu$ is introduced to ensure the coupling 
constant is dimensionless in $d$-dimensions. We have included the abelian gauge 
parameter renormalization for completeness but it will not be discussed here as
the Landau gauge will be chosen for that sector. Also we have provided 
(\ref{zdef}) to highlight that there is a nontrivial aspect to the 
renormalization of QCD in the MAG. When one fixes a gauge the remnant of the 
original gauge symmetry becomes manifest in the Slavnov-Taylor identities via 
the underlying BRST symmetry. These place certain constraints on the 
renormalization constants which must be respected in any computation and 
renormalization scheme. Those identities for the ordinary linear covariant 
gauge fixing are well known and can be established systematically by the 
algebraic renormalization technique \cite{42}. However, when that method is 
applied to the MAG the diagonal ghost and anti-ghost renormalization constants 
are not defined in the canonical way, (\ref{zdef}), \cite{29}. This has been 
checked to three loops in $\MSbar$ in \cite{30}. Therefore, we have to allow 
for this in our definitions. Moreover, to determine the centre ghost 
renormalization at one loop requires a two loop renormalization of a 
{\em vertex} function. A second consequence of the Slavnov-Taylor identities is
that the centre gluon wave function renormalization constant is not an 
independent renormalization. It is related to the coupling constant 
renormalization, \cite{29}, and as such provides an independent check on any 
computation. For completeness the relevant renormalization group functions we 
will consider here in the various MOM schemes are  
\begin{eqnarray}
\gamma_A(a,\alpha) &=& \beta(a,\alpha) \frac{\partial}{\partial a} \ln Z_A ~+~ 
\alpha \gamma_\alpha(a,\alpha) \frac{\partial}{\partial \alpha} \ln Z_A 
\nonumber \\
\gamma_\alpha(a,\alpha) &=& \left[ \beta(a,\alpha) \frac{\partial}{\partial a}
\ln Z_\alpha ~-~ \gamma_A(a,\alpha) \right] \left[ 1 ~-~ \alpha
\frac{\partial}{\partial \alpha} \ln Z_\alpha \right]^{-1} \nonumber \\
\gamma_{A^i}(a,\alpha) &=& \beta(a,\alpha) \frac{\partial}{\partial a} 
\ln Z_{A^i} ~+~ \alpha \gamma_\alpha(a,\alpha) \frac{\partial}{\partial \alpha}
\ln Z_{A^i} \nonumber \\
\gamma_c(a,\alpha) &=& \beta(a,\alpha) \frac{\partial}{\partial a} \ln Z_c ~+~ 
\alpha \gamma_\alpha(a,\alpha) \frac{\partial}{\partial \alpha} \ln Z_c 
\nonumber \\
\gamma_\psi(a,\alpha) &=& \beta(a,\alpha) \frac{\partial}{\partial a} 
\ln Z_\psi ~+~ \alpha \gamma_\alpha(a,\alpha) \frac{\partial}{\partial \alpha} 
\ln Z_\psi 
\end{eqnarray}
where the form of $\gamma_\alpha(a,\alpha)$ is due to the fact that unlike a
linear covariant gauge fixing $Z_\alpha$ is not unity. The quantity $a$ is
defined to be $a$~$=$~$g^2/(16\pi^2)$. Also we have included $\alpha$ 
dependence on the $\beta$-function since in mass dependent renormalization 
schemes such as the MOM cases we consider here the $\beta$-function is gauge 
dependent. 

Having defined the renormalization group functions in relation to the
renormalization constants for a particular scheme we now recall the formalism
which relates the expressions between two different schemes. First we note
that the parameters such as the coupling constant and the gauge parameter are
associated with a scheme and therefore their values differ between schemes.
They are related via their respective renormalization constants. In 
particular\footnote{The second equation corrects an obvious error in the
corresponding relation in \cite{19}.}
\begin{equation}
g_{\MOMis}(\mu) ~=~ \frac{Z_g^{\MSbars}}{Z_g^{\MOMis}} \, 
g_{\MSbars}(\mu) ~~~,~~~
\alpha_{\MOMis}(\mu) ~=~ \frac{Z_A^{\MSbars}Z_\alpha^{\MOMis}}
{Z_A^{\MOMis}Z_\alpha^{\MSbars}} \, \alpha_{\MSbars}(\mu) 
\label{paramdef}
\end{equation}
where we use $\MOMi$ to label a typical MOM scheme and use these as well as the
$\MSbar$ scheme to illustrate the formalism for converting between schemes.
However, it is important to realise that the explicit relation between the
parameters is found recursively. This is because on the right hand side of each
of the equations of (\ref{paramdef}) the $\MOMi$ renormalization constant is a 
function of the parameters in that scheme. Therefore these have to be mapped 
order by order in the perturbative expansion to the reference scheme, which 
will be $\MSbar$ throughout, prior to extracting the parameter relation at a 
particular loop order. Otherwise one would not have a relation between 
parameters which is finite with respect to the regularization. Once the mapping
of the parameters from one scheme to another has been established it is 
possible to define conversion functions for all the renormalization group
functions. These are similar to (\ref{paramdef}) and are given by 
\begin{equation}
C^{\MOMis}_g(a,\alpha) ~=~ \frac{Z_g^{\MOMis}}{Z_g^{\MSbars}} ~~~~,~~~~
C^{\MOMis}_\phi(a,\alpha) ~=~ \frac{Z_\phi^{\MOMis}}{Z_\phi^{\MSbars}}
\end{equation}
where $\phi$ denotes the field associated with the anomalous dimension and the
arguments of the conversion functions are the $\MSbar$ parameters as this is 
the reference scheme. Though for the gauge parameter we define 
\begin{equation}
C^{\MOMis}_\alpha(a,\alpha) ~=~ \frac{Z_\alpha^{\MOMis}Z_A^{\MSbars}}
{Z_\alpha^{\MSbars}Z_A^{\MOMis}}
\end{equation}
as the conversion function. Again the perturbative expansion of each 
conversion function is finite with respect to $\epsilon$ at each order once the
parameter mapping has been applied. Equipped with these then the relations 
between the renormalization group functions in the various schemes are
\begin{eqnarray}
\beta^{\mbox{$\MOMis$}} ( a_{\mbox{$\MOMis$}}, \alpha_{\mbox{$\MOMis$}} ) &=&
\left[ \beta^{\mbox{$\MSbars$}}( a_{\mbox{$\MSbars$}} )
\frac{\partial a_{\mbox{$\MOMis$}}}{\partial a_{\mbox{$\MSbars$}}} \,+\,
\alpha_{\mbox{$\MSbars$}} \gamma^{\mbox{$\MSbars$}}_\alpha
( a_{\mbox{$\MSbars$}}, \alpha_{\mbox{\footnotesize{$\MSbars$}}} )
\frac{\partial a_{\mbox{$\MOMis$}}}{\partial \alpha_{\mbox{$\MSbars$}}}
\right]_{ \MSbars \rightarrow \MOMis } \nonumber \\
\label{betamap}
\end{eqnarray}
and
\begin{eqnarray}
\gamma_\phi^{\MOMis} ( a_{\MOMis}, \alpha_{\MOMis} )
&=& \!\! \! \left[ \gamma_\phi^{\MSbars} \left(a_{\MSbars}\right)
+ \beta^{\MSbars}\left(a_{\MSbars}\right)
\frac{\partial ~}{\partial a_{\MSbars}} \ln C_\phi^{\MOMis}
\left(a_{\MSbars},\alpha_{\MSbars}\right) \right. \nonumber \\
&& \left. +~ \alpha_{\MSbars} \gamma^{\MSbars}_\alpha
\left(a_{\MSbars},\alpha_{\MSbars}\right)
\frac{\partial ~}{\partial \alpha_{\MSbars}}
\ln C_\phi^{\MOMis} \left(a_{\MSbars},\alpha_{\MSbars}\right)
\right]_{ \MSbars \rightarrow \MOMis } \nonumber \\
\label{gammamap}
\end{eqnarray}
where $\phi$ also includes $\alpha$ here now and the subscript $\MSbars 
\rightarrow \MOMis$ indicates there is a mapping of the parameters after the 
evaluation of the quantity. The $\MSbar$ parameters in the square parentheses 
are mapped to those of the $\MOMi$ scheme. We have written (\ref{betamap}) in 
this particular form in order to indicate its derivation originates from the 
renormalization group formalism but the two derivatives can be simply related 
to $C^{\MOMis}_g(a,\alpha)$. From (\ref{betamap}) and (\ref{gammamap}) it is 
clear from examining the $a$ dependence that to deduce the two loop 
renormalization group functions in the $\MOMi$ scheme only the one loop 
conversion functions are needed as the two loop $\MSbar$ renormalization group 
functions are known. In essence the conversion functions derive from the finite
parts of the vertex functions after renormalization which we will deduce as 
part of our computations for each of the $3$-point vertices.  

Given the structure of the trivalent vertices in the Lagrangian and our aim of 
computing in the MOM setup it appears that there are six such possible schemes.
This is in contrast to the linear covariant gauge fixing where there are three 
schemes deriving from the triple gluon, ghost-gluon and quark-gluon vertices. 
However, in the MAG there are only three rather than the potential six MOM 
schemes. This is because the Slavnov-Taylor identity renders the vertices 
involving the centre gluons effectively trivial. The coupling constant 
renormalization constant derived from these vertices is already determined by 
this identity. Thus the three schemes we will focus on are those which are 
completely parallel to those of \cite{18} where the gluon is off-diagonal. 
Given this we recall the computational setup which will be completely parallel 
to \cite{19}. First we decompose each vertex function at the symmetric 
subtraction point into the scalar amplitudes with their associated Lorentz 
tensor basis. Factoring off the overall colour tensors for each vertex function
using  
\begin{eqnarray}
\left. \left\langle A^a_\mu(p) A^b_\nu(q) A^c_\sigma(r)
\right\rangle \right|_{p^2 = q^2 = - \mu^2} &=& f^{abc}
\left. \Sigma^{\mbox{\footnotesize{ggg}}}_{\mu \nu \sigma}(p,q)
\right|_{p^2 = q^2 = - \mu^2} \nonumber \\
\left. \left\langle c^a(p) \bar{c}^b(q) A^c_\sigma(r)
\right\rangle \right|_{p^2 = q^2 = - \mu^2} &=& f^{abc}
\left. \Sigma^{\mbox{\footnotesize{ccg}}}_\sigma(p,q)
\right|_{p^2 = q^2 = - \mu^2} \nonumber \\
\left. \left\langle \psi^i(p) \bar{\psi}^j(q) A^c_\sigma(r)
\right\rangle \right|_{p^2 = q^2 = - \mu^2} &=& T^c_{ij}
\left. \Sigma^{\mbox{\footnotesize{qqg}}}_\sigma(p,q)
\right|_{p^2 = q^2 = - \mu^2}
\end{eqnarray}
then we write 
\begin{eqnarray}
\left. \frac{}{} \Sigma^{\mbox{\footnotesize{ggg}}}_{\mu \nu \sigma}
(p,q) \right|_{p^2 = q^2 = - \mu^2} &=& \sum_{k=1}^{14}
{\cal P}^{\mbox{\footnotesize{ggg}}}_{(k) \, \mu \nu \sigma }(p,q) \,
\Sigma^{\mbox{\footnotesize{ggg}}}_{(k)}(p,q) \nonumber \\
\left. \frac{}{} \Sigma^{\mbox{\footnotesize{ccg}}}_\sigma(p,q)
\right|_{p^2 = q^2 = - \mu^2} &=& \sum_{k=1}^{2}
{\cal P}^{\mbox{\footnotesize{ccg}}}_{(k) \, \sigma }(p,q) \,
\Sigma^{\mbox{\footnotesize{ccg}}}_{(k)}(p,q) \nonumber \\
\left. \frac{}{} \Sigma^{\mbox{\footnotesize{qqg}}}_\sigma(p,q)
\right|_{p^2 = q^2 = - \mu^2} &=& \sum_{k=1}^{6}
{\cal P}^{\mbox{\footnotesize{qqg}}}_{(k) \, \sigma }(p,q) \,
\Sigma^{\mbox{\footnotesize{qqg}}}_{(k)}(p,q) ~.
\end{eqnarray}
Throughout we use $p$ and $q$ as the two independent external momenta which
will be the incoming momenta for the ghost and quark lines in their respective 
cases. The third external momentum is $r$ where
\begin{equation}
r ~=~ -~ p ~-~ q ~.
\end{equation}
The symmetric point is then defined as 
\begin{equation}
p^2 ~=~ q^2 ~=~ r^2 ~=~ -~ \mu^2
\end{equation}
which implies
\begin{equation}
pq ~=~ \frac{1}{2} \mu^2 ~.
\end{equation}
To determine each scalar amplitude within a vertex function we use the same
projection method and tensor basis as \cite{19} where the explicit derivation 
is detailed. The explicit forms of the tensors and projection matrices 
${\cal M}^i_{kl}$, where $i$ denotes the vertex, are given for completeness in 
appendix A. Though we recall that 
\begin{eqnarray}
f^{abc} \Sigma^{\mbox{\footnotesize{ggg}}}_{(k)}(p,q)
&=& {\cal M}^{\mbox{\footnotesize{ggg}}}_{kl} \left(
{\cal P}^{\mbox{\footnotesize{ggg}} \, \mu \nu \sigma}_{(l)}(p,q) \left.
\left\langle A^a_\mu(p) A^b_\nu(q) A^c_\sigma(r)
\right\rangle \right )\right|_{p^2 = q^2 = - \mu^2} \nonumber \\
f^{abc} \Sigma^{\mbox{\footnotesize{ccg}}}_{(k)}(p,q) &=&
{\cal M}^{\mbox{\footnotesize{ccg}}}_{kl} \left(
{\cal P}^{\mbox{\footnotesize{ccg}} \, \sigma}_{(l)}(p,q) \left.
\left\langle c^a(p) \bar{c}^b(q) A^c_\sigma(r)
\right\rangle \right) \right|_{p^2 = q^2 = - \mu^2} \nonumber \\
T^c_{ij} \Sigma^{\mbox{\footnotesize{qqg}}}_{(k)}(p,q) &=&
{\cal M}^{\mbox{\footnotesize{qqg}}}_{kl} \left(
{\cal P}^{\mbox{\footnotesize{qqg}} \, \sigma}_{(l)}(p,q) \left.
\left\langle \psi^i(p) \bar{\psi}^j(q) A^c_\sigma(r)
\right\rangle \right) \right|_{p^2 = q^2 = - \mu^2}
\end{eqnarray}
are the linear combinations for each Lorentz channel. For the quark-gluon
vertex we use the generalized $\gamma$-matrices denoted by
$\Gamma_{(n)}$ and defined by
\begin{equation}
\Gamma_{(n)}^{\mu_1 \ldots \mu_n} ~=~ \gamma^{[\mu_1} \ldots \gamma^{\mu_n]}
\end{equation}
where a factor $1/n!$ is understood in the total antisymmetrization. 
Properties of these matrices have been detailed in \cite{43,44,45,46,47}. In 
this basis which spans the space of $\gamma$-matrices there is a natural 
partition due to 
\begin{equation}
\mbox{tr} \left( \Gamma_{(m)}^{\mu_1 \ldots \mu_m}
\Gamma_{(n)}^{\nu_1 \ldots \nu_n} \right) ~ \propto ~ \delta_{mn}
I^{\mu_1 \ldots \mu_m \nu_1 \ldots \nu_n} 
\end{equation}
which is evident in (\ref{qqgm}). 

Once all the vertices have been decomposed into their Lorentz scalars we have
to reduce the large number of Feynman integrals to a form in which they can be
evaluated. We have chosen to use the Laporta approach, \cite{48}. This method
allows one to construct all the integration by parts identities for a minimal
set of basic topologies. Suitably chosen these cover all possible topologies
which arise in the vertex functions. Once the relations between all the
integrals are known then they can be algebraically solved to a small set of
master graphs. Ordinarily these have to be determined by non-integration by
parts methods. In practical terms the Laporta algorithm has been coded in
several packages. We have used {\sc Reduze}, \cite{49}, which uses {\sc GiNaC},
\cite{50}, and built the necessary database. At one loop there is one basis 
topology. For each vertex function we have used {\sc Qgraf}, \cite{51}, to 
generate all the Feynman graphs and then mapped them on to the basic 
topologies. We have appended colour and Lorentz indices in the initial steps 
too. Throughout we have used {\sc Form}, \cite{52,53}, as the computational 
tool to handle the algebra symbolically. For the triple off-diagonal vertex 
there are $23$ one loop graphs. The ghost-gluon vertex has $16$ graphs and 
there are $5$ for the quark-gluon vertex. Briefly for the $2$-point functions 
we have used {\sc Mincer}, \cite{54,55}, in order to evaluate the small number 
of straightforward graphs. 

Given that we are working in the MAG it is worthwhile recalling some of the
group theory identities which we have had to use, \cite{30}. As the colour 
group has been split into two sectors we have to be careful in implementing 
this symbolically. Useful in this instance is the set facility in {\sc Form} in
order to treat centre and off-diagonal indices separately. The starting point
for deriving any group identities for the split Lie algebra is the original
identities. First, the usual Casimirs are defined by
\begin{equation}
f^{ACD} f^{BCD} ~=~ C_A \delta^{AB} ~~,~~ T^A T^A ~=~ C_F I ~~,~~
\mbox{Tr} \left( T^A T^B \right) ~=~ T_F \delta^{AB} 
\label{Cas}
\end{equation}
where $I$ is the identity. The former gives the non-trivial results 
\begin{equation}
C_A \delta^{ab} ~=~ f^{acd} f^{bcd} ~+~ 2 f^{acj} f^{bcj} ~~,~~ 
C_A \delta^{ij} ~=~ f^{icd} f^{jcd} 
\end{equation}
if one recalls the structure functions can only have at most one centre
index. These imply 
\begin{eqnarray}
f^{iab} f^{iab} &=& \Nda C_A ~~,~~
f^{abc} f^{abc} ~=~ \left[ \Noda - 2 \Nda \right] C_A \nonumber \\
f^{acj} f^{bcj} &=& \frac{\Nda}{\Noda} C_A \delta^{ab} ~~,~~
f^{acd} f^{bcd} ~=~ \frac{[\Noda - 2 \Nda]}{\Noda} C_A \delta^{ab} ~.
\end{eqnarray}
The remaining equations of (\ref{Cas}) give the simple expressions
\begin{equation}
\mbox{Tr} \left( T^a T^b \right) ~=~ T_F \delta^{ab} ~~,~~
\mbox{Tr} \left( T^a T^i \right) ~=~ 0 ~~,~~
\mbox{Tr} \left( T^i T^j \right) ~=~ T_F \delta^{ij}
\end{equation}
as well as 
\begin{equation}
T^i T^i ~=~ \frac{T_F}{\NF} \Nda I ~~,~~ 
T^a T^a ~=~ \left[ C_F ~-~ \frac{T_F}{\NF} \Nda \right] I ~.
\end{equation}
The Jacobi identity 
\begin{equation}
0 ~=~ f^{ABE} f^{CDE} ~+~ f^{BCE} f^{ADE} ~+~ f^{CAE} f^{BDE} 
\end{equation}
provides more results which we needed such as 
\begin{eqnarray}
f^{apq} f^{bpr} f^{cqr} &=& \frac{[\Noda - 3 \Nda]}{2\Noda} C_A f^{abc} ~~,~~
f^{apq} f^{bpi} f^{cqi} ~=~ \frac{\Nda}{2\Noda} C_A f^{abc}
\nonumber \\
f^{ipq} f^{bpr} f^{cqr} &=& \frac{[\Noda - 2 \Nda]}{2\Noda} C_A f^{ibc} ~~,~~
f^{ipq} f^{bpj} f^{cqj} ~=~ \frac{\Nda}{\Noda} C_A f^{ibc} ~.
\end{eqnarray}
A useful relation between dimensions is
\begin{equation}
C_F \NF ~=~ \left( \Nda ~+~ \Noda \right) T_F
\end{equation}
which is required usually for simplifying algebra from the quark sector. These 
basic results and others have been coded within a {\sc Form} module and applied
prior to the integrals being mapped to the basic topologies. This is because as
was noted in \cite{30} the group theory for some graphs is zero. Hence in such 
cases there is no need for a calculation to be performed. 

\sect{$\MSbar$ scheme.}

As a preliminary to the MOM scheme computations we first record the results for
the amplitudes in the $\MSbar$ scheme. This is the basic reference scheme.
Indeed to deduce the two loop MOM scheme renormalization group functions using
the conversion functions, the two loop $\MSbar$ results are necessary.
Therefore, for completeness we note that these are\footnote{Electronic versions
of all the MAG renormalization group functions, conversion functions and the 
$\MSbar$ amplitudes for each of the three vertices and the vertex associated 
with its $\MOMi$ scheme are available in the attached data file.}, 
\cite{29,30},
\begin{eqnarray}
\gamma_A(a) &=& \frac{1}{6\Noda} \left[ \Noda \left( ( 3 \alpha - 13 ) C_A
+ 8 T_F \Nf \right) + \Nda ( - 3 \alpha + 9 ) C_A \right] a \nonumber \\
&& +~ \frac{1}{48{\Noda}^2} \left[ {\Noda}^2 \left( ( 6 \alpha^2 + 66 \alpha
- 354 ) C_A^2 + 240 C_A T_F \Nf + 192 C_F T_F \Nf \right) \right. \nonumber \\
&& \left. ~~~~~~~~~~~~~+~ \Noda \Nda \left( ( 3 \alpha^2 + 210 \alpha
+ 331 ) C_A^2 - 80 C_A T_F \Nf \right) \right. \nonumber \\
&& \left. ~~~~~~~~~~~~~+~ {\Nda}^2 \left( ( 15 \alpha^2 - 6 \alpha - 33 )
C_A^2 \right) \right] a^2 ~+~ O(a^3) \nonumber \\
\gamma_\alpha(a) &=& \frac{1}{12 \alpha\Noda} \left[ \Noda \left(
( -~ 3 \alpha^2 + 26 \alpha ) C_A - 16 \alpha T_F \Nf \right)
+ \Nda ( -~ 6 \alpha^2 - 36 \alpha - 36 ) C_A \right] a
\nonumber \\
&& +~ \frac{1}{48 \alpha {\Noda}^2} \left[ {\Noda}^2 \left( ( -~ 3 \alpha^3
- 51 \alpha^2 + 354 \alpha ) C_A^2 - 240 \alpha C_A T_F \Nf - 192 \alpha
C_F T_F \Nf \right) \right. \nonumber \\
&& \left. ~~~~~~~~~~~~~~~+~ \Noda \Nda \left( ( -\, 27 \alpha^3 - 339 \alpha^2
- 647 \alpha - 928 ) C_A^2 \right. \right. \nonumber \\
&& \left. \left. ~~~~~~~~~~~~~~~~~~~~~~~~~~~~~+~ ( 160 \alpha + 512 ) 
C_A T_F \Nf \right) \right.  \nonumber \\
&& \left. ~~~~~~~~~~~~~~~+~ {\Nda}^2 ( -~ 30 \alpha^3 - 366 \alpha^2 
+ 294 \alpha + 2016 ) C_A^2 \right] a^2 ~+~ O(a^3) \nonumber \\
\gamma_{A^i}(a) &=& \frac{1}{3} \left[ 4 T_F \Nf - 11 C_A \right] a
\nonumber \\
&& +~ \frac{1}{3} \left[ -~ 34 C_A^2 + 20 C_A T_F \Nf + 12 C_F T_F \Nf \right]
a^2 ~+~ O(a^3) \nonumber \\
\gamma_c(a) &=& \frac{1}{4\Noda} \left[ \Noda ( \alpha - 3 ) C_A
+ \Nda ( - 2 \alpha - 6 ) C_A \right] a \nonumber \\
&& +~ \frac{1}{96 {\Noda}^2} \left[ {\Noda}^2 \left( ( 6 \alpha^2 - 6 \alpha
- 190 ) C_A^2 + 80 C_A T_F \Nf \right) \right. \nonumber \\
&& \left. ~~~~~~~~~~~~~~+~ \Noda \Nda \left( ( -~ 42 \alpha^2 - 126 \alpha
- 347 ) C_A^2 + 160 C_A T_F \Nf \right) \right. \nonumber \\
&& \left. ~~~~~~~~~~~~~~+~ {\Nda}^2 ( 12 \alpha^2 - 588 \alpha + 510 )
C_A^2 \right] a^2 ~+~ O(a^3) \nonumber \\
\gamma_{c^i}(a) &=& \frac{1}{4 \Noda} \left[ \Noda ( - \alpha - 3 ) C_A
+ \Nda ( - 2 \alpha - 6 ) C_A \right] a \nonumber \\
&& +~ \frac{1}{96 {\Noda}^2} \left[ {\Noda}^2 \left( ( -~ 6 \alpha^2
- 66 \alpha - 190 ) C_A^2 + 80 C_A T_F \Nf \right) \right. \nonumber \\
&& \left. ~~~~~~~~~~~~~~+~ \Noda \Nda \left( ( -~ 54 \alpha^2 - 354 \alpha
- 323 ) C_A^2 + 160 C_A T_F \Nf \right) \right. \nonumber \\
&& \left. ~~~~~~~~~~~~~~+~ {\Nda}^2 ( -~ 60 \alpha^2 - 372 \alpha + 510 ) C_A^2
\right] a^2 ~+~ O(a^3) \nonumber \\
\gamma_\psi(a) &=& \frac{\alpha \Noda T_F}{\NF} a \nonumber \\
&& +~ \frac{1}{4\NF} \left[ ( -~ \alpha^2 + 22 \alpha + 23 ) C_A C_F \NF
+ ( \alpha^2 - 14 \alpha + 2 ) \Noda C_A T_F \right. \nonumber \\
&& \left. ~~~~~~~~~~~-~ 6 C_F^2 \NF - 8 C_F \Nf T_F \NF \right] a^2 ~+~ 
O(a^3) ~.
\end{eqnarray}
Though the three loop results are also available, \cite{30}. 

Next the full one loop amplitudes for each of the three vertex functions which
we have calculated here in $\MSbar$ are 
\begin{eqnarray}
\left. \Sigma^{\mbox{\footnotesize{ggg}}}_{(1)}(p,q) \right|_{\MSbars} &=&
\left. \Sigma^{\mbox{\footnotesize{ggg}}}_{(2)}(p,q) \right|_{\MSbars} ~=~
-~ \frac{1}{2} \left. \Sigma^{\mbox{\footnotesize{ggg}}}_{(3)}(p,q)
\right|_{\MSbars} ~=~
-~ \left. \Sigma^{\mbox{\footnotesize{ggg}}}_{(4)}(p,q) \right|_{\MSbars}
\nonumber \\
&=& \frac{1}{2} \left. \Sigma^{\mbox{\footnotesize{ggg}}}_{(5)}(p,q)
\right|_{\MSbars} ~=~
-~ \left. \Sigma^{\mbox{\footnotesize{ggg}}}_{(6)}(p,q) \right|_{\MSbars}
\nonumber \\
&=& 1 + \left[
-~ 72 \psi^\prime(\third) \alpha^2 C_A {\Nda}
+ 36 \psi^\prime(\third) \alpha^2 C_A {\Noda}
+ 90 \psi^\prime(\third) \alpha C_A {\Nda}
\right. \nonumber \\
&& \left. ~~~~~
-~ 162 \psi^\prime(\third) \alpha C_A {\Noda}
- 702 \psi^\prime(\third) C_A {\Nda}
+ 138 \psi^\prime(\third) C_A {\Noda}
\right. \nonumber \\
&& \left. ~~~~~
-~ 384 \psi^\prime(\third) \Nf {\Noda} T_F
- 81 \alpha^3 C_A {\Nda}
+ 27 \alpha^3 C_A {\Noda}
+ 48 \pi^2 \alpha^2 C_A {\Nda}
\right. \nonumber \\
&& \left. ~~~~~
+~ 810 \alpha^2 C_A {\Nda}
- 24 \pi^2 \alpha^2 C_A {\Noda}
- 405 \alpha^2 C_A {\Noda}
- 60 \pi^2 \alpha C_A {\Nda}
\right. \nonumber \\
&& \left. ~~~~~
+~ 243 \alpha C_A {\Nda}
+ 108 \pi^2 \alpha C_A {\Noda}
- 243 \alpha C_A {\Noda}
+ 468 \pi^2 C_A {\Nda}
\right. \nonumber \\
&& \left. ~~~~~
+~ 2916 C_A {\Nda}
- 92 \pi^2 C_A {\Noda}
- 243 C_A {\Noda}
+ 256 \pi^2 \Nf {\Noda} T_F
\right. \nonumber \\
&& \left. ~~~~~
+~ 1296 \Nf {\Noda} T_F \right] \frac{a}{648 {\Noda}} ~+~ O(a^2) \nonumber \\
\left. \Sigma^{\mbox{\footnotesize{ggg}}}_{(7)}(p,q) \right|_{\MSbars} &=&
2 \left. \Sigma^{\mbox{\footnotesize{ggg}}}_{(9)}(p,q) \right|_{\MSbars} ~=~
-~ 2 \left. \Sigma^{\mbox{\footnotesize{ggg}}}_{(11)}(p,q)
\right|_{\MSbars} ~=~
-~ \left. \Sigma^{\mbox{\footnotesize{ggg}}}_{(14)}(p,q) \right|_{\MSbars}
\nonumber \\
&=&  \left[ 
108 \psi^\prime(\third) \alpha^5 C_A {\Nda}
- 36 \psi^\prime(\third) \alpha^5 C_A {\Noda}
- 324 \psi^\prime(\third) \alpha^4 C_A {\Nda}
\right. \nonumber \\
&& \left. \,
+~ 162 \psi^\prime(\third) \alpha^4 C_A {\Noda}
+ 324 \psi^\prime(\third) \alpha^3 C_A {\Nda}
- 108 \psi^\prime(\third) \alpha^3 C_A {\Noda}
\right. \nonumber \\
&& \left. \,
+~ 1296 \psi^\prime(\third) \alpha^2 C_A {\Nda}
- 456 \psi^\prime(\third) \alpha^2 C_A {\Noda}
+ 768 \psi^\prime(\third) \alpha^2 \Nf {\Noda} T_F
\right. \nonumber \\
&& \left. \,
+~ 216 \psi^\prime(\third) \alpha C_A {\Nda}
+ 270 \psi^\prime(\third) C_A {\Nda}
- 72 \pi^2 \alpha^5 C_A {\Nda}
- 324 \alpha^5 C_A {\Nda}
\right. \nonumber \\
&& \left. \,
+~ 24 \pi^2 \alpha^5 C_A {\Noda}
+ 108 \alpha^5 C_A {\Noda}
+ 216 \pi^2 \alpha^4 C_A {\Nda}
+ 810 \alpha^4 C_A {\Nda}
\right. \nonumber \\
&& \left. \,
-~ 108 \pi^2 \alpha^4 C_A {\Noda}
- 405 \alpha^4 C_A {\Noda}
- 216 \pi^2 \alpha^3 C_A {\Nda}
- 1377 \alpha^3 C_A {\Nda}
\right. \nonumber \\
&& \left. \,
+~ 72 \pi^2 \alpha^3 C_A {\Noda}
+ 1458 \alpha^3 C_A {\Noda}
- 864 \pi^2 \alpha^2 C_A {\Nda}
+ 891 \alpha^2 C_A {\Nda}
\right. \nonumber \\
&& \left. \,
+~ 304 \pi^2 \alpha^2 C_A {\Noda}
- 873 \alpha^2 C_A {\Noda}
- 512 \pi^2 \alpha^2 \Nf {\Noda} T_F
- 576 \alpha^2 \Nf {\Noda} T_F
\right. \nonumber \\
&& \left. \,
-~ 144 \pi^2 \alpha C_A {\Nda}
- 243 \alpha C_A {\Nda}
- 180 \pi^2 C_A {\Nda}
+ 243 C_A {\Nda} \right] \frac{a}{972 \alpha^2 {\Noda}} \nonumber \\
&& +~ O(a^2) \nonumber \\
\left. \Sigma^{\mbox{\footnotesize{ggg}}}_{(8)}(p,q) \right|_{\MSbars} &=&
-~ \left. \Sigma^{\mbox{\footnotesize{ggg}}}_{(13)}(p,q) \right|_{\MSbars}
\nonumber \\
&=& \left[ 
-~ 108 \psi^\prime(\third) \alpha^5 C_A {\Nda}
+ 36 \psi^\prime(\third) \alpha^5 C_A {\Noda}
+ 540 \psi^\prime(\third) \alpha^4 C_A {\Nda}
\right. \nonumber \\
&& \left. \,
-~ 270 \psi^\prime(\third) \alpha^4 C_A {\Noda}
- 270 \psi^\prime(\third) \alpha^3 C_A {\Nda}
+ 378 \psi^\prime(\third) \alpha^3 C_A {\Noda}
\right. \nonumber \\
&& \left. \,
+~ 1242 \psi^\prime(\third) \alpha^2 C_A {\Nda}
- 390 \psi^\prime(\third) \alpha^2 C_A {\Noda}
+ 384 \psi^\prime(\third) \alpha^2 \Nf {\Noda} T_F
\right. \nonumber \\
&& \left. \,
+~ 216 \psi^\prime(\third) \alpha C_A {\Nda}
+ 270 \psi^\prime(\third) C_A {\Nda}
+ 72 \pi^2 \alpha^5 C_A {\Nda}
\right. \nonumber \\
&& \left. \,
+~ 567 \alpha^5 C_A {\Nda}
- 24 \pi^2 \alpha^5 C_A {\Noda}
- 189 \alpha^5 C_A {\Noda}
- 360 \pi^2 \alpha^4 C_A {\Nda}
\right. \nonumber \\
&& \left. \,
-~ 2268 \alpha^4 C_A {\Nda}
+ 180 \pi^2 \alpha^4 C_A {\Noda}
+ 1134 \alpha^4 C_A {\Noda}
+ 180 \pi^2 \alpha^3 C_A {\Nda}
\right. \nonumber \\
&& \left. \,
+~ 648 \alpha^3 C_A {\Nda}
- 252 \pi^2 \alpha^3 C_A {\Noda}
- 243 \alpha^3 C_A {\Noda}
- 828 \pi^2 \alpha^2 C_A {\Nda}
\right. \nonumber \\
&& \left. \,
+~ 1053 \alpha^2 C_A {\Nda}
+ 260 \pi^2 \alpha^2 C_A {\Noda}
- 1206 \alpha^2 C_A {\Noda}
- 256 \pi^2 \alpha^2 \Nf {\Noda} T_F
\right. \nonumber \\
&& \left. \,
+~ 1008 \alpha^2 \Nf {\Noda} T_F
- 144 \pi^2 \alpha C_A {\Nda}
- 243 \alpha C_A {\Nda}
- 180 \pi^2 C_A {\Nda}
\right. \nonumber \\
&& \left. \,
+~ 243 C_A {\Nda} \right] \frac{a}{972 \alpha^2 {\Noda}} ~+~ O(a^2) 
\nonumber \\
\left. \Sigma^{\mbox{\footnotesize{ggg}}}_{(10)}(p,q) \right|_{\MSbars} &=&
-~ \left. \Sigma^{\mbox{\footnotesize{ggg}}}_{(12)}(p,q) \right|_{\MSbars}
\nonumber \\
&=& \left[ 
216 \psi^\prime(\third) \alpha^3 C_A {\Nda}
- 72 \psi^\prime(\third) \alpha^3 C_A {\Noda}
- 864 \psi^\prime(\third) \alpha^2 C_A {\Nda}
\right. \nonumber \\
&& \left. \,
+~ 432 \psi^\prime(\third) \alpha^2 C_A {\Noda}
+ 594 \psi^\prime(\third) \alpha C_A {\Nda}
- 486 \psi^\prime(\third) \alpha C_A {\Noda}
\right. \nonumber \\
&& \left. \,
+~ 54 \psi^\prime(\third) C_A {\Nda}
- 66 \psi^\prime(\third) C_A {\Noda}
+ 384 \psi^\prime(\third) \Nf {\Noda} T_F
\right. \nonumber \\
&& \left. \,
-~ 144 \pi^2 \alpha^3 C_A {\Nda}
- 891 \alpha^3 C_A {\Nda}
+ 48 \pi^2 \alpha^3 C_A {\Noda}
+ 297 \alpha^3 C_A {\Noda}
\right. \nonumber \\
&& \left. \,
+~ 576 \pi^2 \alpha^2 C_A {\Nda}
+ 3078 \alpha^2 C_A {\Nda}
- 288 \pi^2 \alpha^2 C_A {\Noda}
- 1539 \alpha^2 C_A {\Noda}
\right. \nonumber \\
&& \left. \,
-~ 396 \pi^2 \alpha C_A {\Nda}
- 2025 \alpha C_A {\Nda}
+ 324 \pi^2 \alpha C_A {\Noda}
+ 1701 \alpha C_A {\Noda}
\right. \nonumber \\
&& \left. \,
-~ 36 \pi^2 C_A {\Nda}
- 162 C_A {\Nda}
+ 44 \pi^2 C_A {\Noda}
+ 333 C_A {\Noda}
\right. \nonumber \\
&& \left. \,
-~ 256 \pi^2 \Nf {\Noda} T_F
- 1584 \Nf {\Noda} T_F \right] \frac{a}{972 {\Noda}} ~+~ O(a^2) 
\end{eqnarray}
for the triple gluon vertex. Those for the other two vertices are 
\begin{eqnarray}
\left. \Sigma^{\mbox{\footnotesize{ccg}}}_{(1)}(p,q) \right|_{\MSbars} &=&
-~ \left. \Sigma^{\mbox{\footnotesize{ccg}}}_{(2)}(p,q) \right|_{\MSbars}
\nonumber \\
&=& \frac{1}{2} + \left[
18 \psi^\prime(\third) \alpha {\Nda}
- 6 \psi^\prime(\third) \alpha {\Noda}
- 33 \psi^\prime(\third) {\Nda}
+ 15 \psi^\prime(\third) {\Noda}
- 12 \alpha {\Nda} \pi^2
\right. \nonumber \\
&& \left. ~~~~~~
- 27 \alpha {\Nda}
+ 4 \alpha {\Noda} \pi^2
+ 27 \alpha {\Noda}
+ 22 {\Nda} \pi^2
+ 27 {\Nda}
- 10 {\Noda} \pi^2
\right. \nonumber \\
&& \left. ~~~~~~
+ 81 {\Noda} \right] \frac{C_A a}{216 {\Noda}} ~+~ O(a^2)
\end{eqnarray}
and
\begin{eqnarray}
\left. \Sigma^{\mbox{\footnotesize{qqg}}}_{(1)}(p,q) \right|_{\MSbars} &=&
-~ 1 + \left[ 6 \psi^\prime(\third) \alpha^2 C_A \NF {\Nda}
- 3 \psi^\prime(\third) \alpha^2 C_A \NF {\Noda}
- 12 \psi^\prime(\third) \alpha C_A \NF {\Nda}
\right. \nonumber \\
&& \left. ~~~~~~~~~
+~ 12 \psi^\prime(\third) \alpha C_A \NF {\Noda}
+ 48 \psi^\prime(\third) \alpha {\Noda}^2 T_F
+ 30 \psi^\prime(\third) C_A \NF {\Nda}
\right. \nonumber \\
&& \left. ~~~~~~~~~
+~ 39 \psi^\prime(\third) C_A \NF {\Noda}
- 24 \psi^\prime(\third) C_F \NF {\Noda}
- 4 \pi^2 \alpha^2 C_A \NF {\Nda}
\right. \nonumber \\
&& \left. ~~~~~~~~~
-~ 54 \alpha^2 C_A \NF {\Nda}
+ 2 \pi^2 \alpha^2 C_A \NF {\Noda}
+ 27 \alpha^2 C_A \NF {\Noda}
\right. \nonumber \\
&& \left. ~~~~~~~~~
+~ 8 \pi^2 \alpha C_A \NF {\Nda}
- 8 \pi^2 \alpha C_A \NF {\Noda}
- 32 \pi^2 \alpha {\Noda}^2 T_F
\right. \nonumber \\
&& \left. ~~~~~~~~~
-~ 216 \alpha {\Noda}^2 T_F
- 20 \pi^2 C_A \NF {\Nda}
- 162 C_A \NF {\Nda}
- 26 \pi^2 C_A \NF {\Noda}
\right. \nonumber \\
&& \left. ~~~~~~~~~
-~ 351 C_A \NF {\Noda}
+ 16 \pi^2 C_F \NF {\Noda}
+ 216 C_F \NF {\Noda} \right] \frac{a}{108 \NF {\Noda}} \nonumber \\
&& +~ O(a^2) \nonumber \\
\left. \Sigma^{\mbox{\footnotesize{qqg}}}_{(2)}(p,q) \right|_{\MSbars} &=&
\left. \Sigma^{\mbox{\footnotesize{qqg}}}_{(5)}(p,q) \right|_{\MSbars}
\nonumber \\
&=& \left[ 6 \psi^\prime(\third) \alpha^2 C_A \NF {\Nda}
- 3 \psi^\prime(\third) \alpha^2 C_A \NF {\Noda}
+ 24 \psi^\prime(\third) \alpha {\Noda}^2 T_F
\right. \nonumber \\
&& \left. \, 
+~ 6 \psi^\prime(\third) C_A \NF {\Nda}
+ 15 \psi^\prime(\third) C_A \NF {\Noda}
- 24 \psi^\prime(\third) C_F \NF {\Noda}
\right. \nonumber \\
&& \left. \, 
-~ 4 \pi^2 \alpha^2 C_A \NF {\Nda}
- 36 \alpha^2 C_A \NF {\Nda}
+ 2 \pi^2 \alpha^2 C_A \NF {\Noda}
+ 18 \alpha^2 C_A \NF {\Noda}
\right. \nonumber \\
&& \left. \, 
-~ 36 \alpha C_A \NF {\Nda}
+ 36 \alpha C_A \NF {\Noda}
- 16 \pi^2 \alpha {\Noda}^2 T_F
- 72 \alpha {\Noda}^2 T_F
\right. \nonumber \\
&& \left. \, 
-~ 4 \pi^2 C_A \NF {\Nda}
+ 36 C_A \NF {\Nda}
- 10 \pi^2 C_A \NF {\Noda}
- 126 C_A \NF {\Noda}
\right. \nonumber \\
&& \left. \, 
+~ 16 \pi^2 C_F \NF {\Noda}
+ 144 C_F \NF {\Noda} \right] \frac{a}{54 \NF {\Noda}} ~+~ O(a^2)
\nonumber \\
\left. \Sigma^{\mbox{\footnotesize{qqg}}}_{(3)}(p,q) \right|_{\MSbars} &=&
\left. \Sigma^{\mbox{\footnotesize{qqg}}}_{(4)}(p,q) \right|_{\MSbars}
\nonumber \\
&=& \left[ 6 \psi^\prime(\third) \alpha C_A \NF {\Nda}
-~ 6 \psi^\prime(\third) \alpha C_A \NF {\Noda}
+ 24 \psi^\prime(\third) \alpha {\Noda}^2 T_F
\right. \nonumber \\
&& \left. \, 
+~ 6 \psi^\prime(\third) C_A \NF {\Nda}
+ 6 \psi^\prime(\third) C_A \NF {\Noda}
- 18 \alpha^2 C_A \NF {\Nda}
+ 9 \alpha^2 C_A \NF {\Noda}
\right. \nonumber \\
&& \left. \, 
-~ 4 \pi^2 \alpha C_A \NF {\Nda}
- 45 \alpha C_A \NF {\Nda}
+ 4 \pi^2 \alpha C_A \NF {\Noda}
+ 45 \alpha C_A \NF {\Noda}
\right. \nonumber \\
&& \left. \, 
-~ 16 \pi^2 \alpha {\Noda}^2 T_F
- 36 \alpha {\Noda}^2 T_F
- 4 \pi^2 C_A \NF {\Nda}
+ 45 C_A \NF {\Nda}
\right. \nonumber \\
&& \left. \, 
-~ 4 \pi^2 C_A \NF {\Noda}
- 90 C_A \NF {\Noda}
+ 72 C_F \NF {\Noda} \right] \frac{a}{54 \NF {\Noda}} ~+~ O(a^2) \nonumber \\
\left. \Sigma^{\mbox{\footnotesize{qqg}}}_{(6)}(p,q) \right|_{\MSbars} &=&
\left[ - 6 \psi^\prime(\third) \alpha^2 C_A {\Nda}
+ 3 \psi^\prime(\third) \alpha^2 C_A {\Noda}
- 12 \psi^\prime(\third) \alpha C_A {\Nda}
+ 12 \psi^\prime(\third) \alpha C_A {\Noda}
\right. \nonumber \\
&& \left. \, 
-~ 6 \psi^\prime(\third) C_A {\Nda}
+ 33 \psi^\prime(\third) C_A {\Noda}
- 24 \psi^\prime(\third) C_F {\Noda}
+ 4 \pi^2 \alpha^2 C_A {\Nda}
\right. \nonumber \\
&& \left. \, 
-~ 2 \pi^2 \alpha^2 C_A {\Noda}
+ 8 \pi^2 \alpha C_A {\Nda}
- 8 \pi^2 \alpha C_A {\Noda}
+ 4 \pi^2 C_A {\Nda}
- 22 \pi^2 C_A {\Noda}
\right. \nonumber \\
&& \left. \, 
+~ 16 \pi^2 C_F {\Noda} \right] \frac{a}{54 {\Noda}} ~+~ O(a^2) 
\end{eqnarray}
where $\psi(z)$ is the derivative of the logarithm of the Euler 
$\Gamma$-function and $\zeta(z)$ is the Riemann zeta function. In the course of
deriving these we have verified that the one loop $\MSbar$ anomalous dimensions
are in agreement with those found in \cite{29,30}. There the vertex functions 
were renormalized at the asymmetric point where the momentum of one of the 
gluon lines for the triple gluon vertex, and the gluon line in the remaining 
vertices was nullified. While this is an exceptional momentum configuration it 
is still possible to extract the $\MSbar$ renormalization constants. In 
changing the subtraction point to the symmetric one the same $\MSbar$ 
renormalization correctly emerged. It is worth noting that essentially the 
contributions to the Lorentz channels containing the poles in $\epsilon$ will 
form the basis for the MOM renormalization. One minor check on the expressions 
is that the correct symmetry structure for each vertex emerged. In other words 
the relations between the various amplitudes for the triple off-diagonal gluon 
vertex, for instance, are consistent with expectations based on \cite{19}. In 
defining the basis of Lorentz tensors we made no assumptions about the
structure which should be present.

\sect{$\MOMggg$ scheme.}

Having discussed the structure of the $3$-point vertices in the $\MSbar$ scheme
at one loop in detail we can now renormalize in each of the MOM schemes defined
by the same vertices. Given that the method and results for each of the 
$\MOMggg$, $\MOMh$ and $\MOMq$ schemes are all effectively the same we focus on
the former and present the full analytic results of the amplitudes for the
vertex defining each scheme. For the other two cases we give condensed versions
in the subsequent sections as the full results are in the data file. With the 
finite parts of the Green's functions being available we define the $\MOMggg$ 
scheme in the MAG in the same way as in QCD, \cite{18}, by ensuring that after 
renormalization there are no $O(a)$ corrections to the Lorentz channels 
containing the divergences in $\epsilon$. In other words for the first six 
amplitudes there are no $O(a)$ parts at the symmetric point but the remaining 
eight amplitudes can have $O(a)$ contributions. Given this and the $\MSbar$ 
results we find that the mappings of the parameters between the schemes are  
\begin{eqnarray}
a_{\MOMgggs} &=&
a + \left[ -~ 72 \psi^\prime(\third) \alpha^2 C_A {\Nda}
+ 36 \psi^\prime(\third) \alpha^2 C_A {\Noda}
+ 90 \psi^\prime(\third) \alpha C_A {\Nda}
\right. \nonumber \\
&& \left. ~~~~~\,
- 162 \psi^\prime(\third) \alpha C_A {\Noda}
- 702 \psi^\prime(\third) C_A {\Nda}
+ 138 \psi^\prime(\third) C_A {\Noda}
- 384 \psi^\prime(\third) \Nf {\Noda} T_F
\right. \nonumber \\
&& \left. ~~~~~\,
- 81 \alpha^3 C_A {\Nda}
+ 27 \alpha^3 C_A {\Noda}
+ 48 \pi^2 \alpha^2 C_A {\Nda}
+ 324 \alpha^2 C_A {\Nda}
\right. \nonumber \\
&& \left. ~~~~~\,
- 24 \pi^2 \alpha^2 C_A {\Noda}
- 162 \alpha^2 C_A {\Noda}
- 60 \pi^2 \alpha C_A {\Nda}
- 243 \alpha C_A {\Nda}
\right. \nonumber \\
&& \left. ~~~~~\,
+ 108 \pi^2 \alpha C_A {\Noda}
+ 243 \alpha C_A {\Noda}
+ 468 \pi^2 C_A {\Nda}
- 92 \pi^2 C_A {\Noda}
\right. \nonumber \\
&& \left. ~~~~~\,
+ 2376 C_A {\Noda}
+ 256 \pi^2 \Nf {\Noda} T_F
- 864 \Nf {\Noda} T_F \right] \frac{a^2}{324 {\Noda}} ~+~ O(a^3) \nonumber \\
\alpha_{\MOMgggs} &=&
\alpha + \left[ 18 \alpha^3 C_A {\Nda} 
- 9 \alpha^3 C_A {\Noda} 
+ 54 \alpha^2 C_A {\Nda} 
- 36 \alpha^2 C_A {\Noda} 
+ 234 \alpha C_A {\Nda} 
\right. \nonumber \\
&& \left. ~~~~~~
- 97 \alpha C_A {\Noda} 
+ 80 \alpha \Nf {\Noda} T_F 
+ 90 C_A {\Nda} \right] \frac{a}{36 {\Noda}} ~+~ O(a^2) ~. 
\end{eqnarray}
Given the nature of the one loop $2$-point functions it transpires that the
gauge parameter mapping is the same for all schemes. This is because the effect
the scheme choice makes on the renormalization of the gauge parameter does not
occur until two loops. The same comment applies to the conversion functions for
the field renormalizations. Therefore, in order to construct the two loop
renormalization group functions we need only record the conversion function for
the coupling constants. For $\MOMggg$ we have 
\begin{eqnarray}
C^{\MOMgggs}_g(a,\alpha) &=& 
1 + \left[ 72 \psi^\prime(\third) \alpha^2 C_A {\Nda}
- 36 \psi^\prime(\third) \alpha^2 C_A {\Noda}
- 90 \psi^\prime(\third) \alpha C_A {\Nda}
\right. \nonumber \\
&& \left. ~~~~~\,
+ 162 \psi^\prime(\third) \alpha C_A {\Noda}
+ 702 \psi^\prime(\third) C_A {\Nda}
- 138 \psi^\prime(\third) C_A {\Noda}
\right. \nonumber \\
&& \left. ~~~~~\,
+ 384 \psi^\prime(\third) \Nf {\Noda} T_F
+ 81 \alpha^3 C_A {\Nda}
- 27 \alpha^3 C_A {\Noda}
- 48 \alpha^2 C_A {\Nda} \pi^2
\right. \nonumber \\
&& \left. ~~~~~\,
- 324 \alpha^2 C_A {\Nda}
+ 24 \alpha^2 C_A {\Noda} \pi^2
+ 162 \alpha^2 C_A {\Noda}
+ 60 \alpha C_A {\Nda} \pi^2
\right. \nonumber \\
&& \left. ~~~~~\,
+ 243 \alpha C_A {\Nda}
- 108 \alpha C_A {\Noda} \pi^2
- 243 \alpha C_A {\Noda}
- 468 C_A {\Nda} \pi^2
\right. \nonumber \\
&& \left. ~~~~~\,
+ 92 C_A {\Noda} \pi^2
- 2376 C_A {\Noda}
- 256 \Nf {\Noda} \pi^2 T_F
+ 864 \Nf {\Noda} T_F \right] \frac{a}{648 {\Noda}} \nonumber \\
&& +~ O(a^2) ~.
\end{eqnarray}
For the other conversion functions we do not label them with the scheme but 
note that like $C^{\MOMgggs}_g(a,\alpha)$ the variables on the left hand side
are the $\MSbar$ ones as is our convention. Thus we have   
\begin{eqnarray}
C_A(a,\alpha) &=& 
1 + \left[ -~ 18 \alpha^2 C_A {\Nda} + 9 \alpha^2 C_A {\Noda} 
- 18 \alpha C_A {\Nda} + 18 \alpha C_A {\Noda} 
- 108 C_A {\Nda} + 97 C_A {\Noda} 
\right. \nonumber \\
&& \left. ~~~~~\,
- 80 \Nf {\Noda} T_F \right] 
\frac{a}{36 {\Noda}} ~+~ O(a^2) \nonumber \\
C_c(a,\alpha) &=& 
1 + C_A \left[ 2 {\Nda} + {\Noda} \right] \frac{a}{{\Noda}} ~+~ O(a^2) 
\nonumber \\
C_\psi(a,\alpha) &=& 
1 - \frac{\alpha {\Noda} T_F a}{\NF} ~+~ O(a^2) ~. 
\end{eqnarray}
Having determined the conversion functions it is straightforward to apply the
renormalization group formalism to construct the two loop $\MOMggg$
renormalization group functions. For the $\beta$-function we find
\begin{eqnarray}
\beta^{\MOMgggs}(a,\alpha) &=&
[ - 11 C_A + 4 \Nf T_F ] \frac{a^2}{3} \nonumber \\
&& + \left[ 288 \psi^\prime(\third) \alpha^3 C_A^2 {\Nda}^2
- 72 \psi^\prime(\third) \alpha^3 C_A^2 {\Noda}^2
+ 1548 \psi^\prime(\third) \alpha^2 C_A^2 {\Nda}^2
\right. \nonumber \\
&& \left. ~~~
- 1878 \psi^\prime(\third) \alpha^2 C_A^2 {\Nda} {\Noda}
+ 786 \psi^\prime(\third) \alpha^2 C_A^2 {\Noda}^2
\right. \nonumber \\
&& \left. ~~~
+ 768 \psi^\prime(\third) \alpha^2 C_A {\Nda} \Nf {\Noda} T_F
- 384 \psi^\prime(\third) \alpha^2 C_A \Nf {\Noda}^2 T_F
\right. \nonumber \\
&& \left. ~~~
+ 648 \psi^\prime(\third) \alpha C_A^2 {\Nda}^2
+ 1860 \psi^\prime(\third) \alpha C_A^2 {\Nda} {\Noda}
\right. \nonumber \\
&& \left. ~~~
- 1404 \psi^\prime(\third) \alpha C_A^2 {\Noda}^2
- 480 \psi^\prime(\third) \alpha C_A {\Nda} \Nf {\Noda} T_F
\right. \nonumber \\
&& \left. ~~~
+ 864 \psi^\prime(\third) \alpha C_A \Nf {\Noda}^2 T_F
- 1080 \psi^\prime(\third) C_A^2 {\Nda}^2
\right. \nonumber \\
&& \left. ~~~
+ 1944 \psi^\prime(\third) C_A^2 {\Nda} {\Noda}
+ 486 \alpha^4 C_A^2 {\Nda}^2
+ 81 \alpha^4 C_A^2 {\Nda} {\Noda}
\right. \nonumber \\
&& \left. ~~~
- 81 \alpha^4 C_A^2 {\Noda}^2
- 192 \pi^2 \alpha^3 C_A^2 {\Nda}^2 
+ 1620 \alpha^3 C_A^2 {\Nda}^2
\right. \nonumber \\
&& \left. ~~~
- 3078 \alpha^3 C_A^2 {\Nda} {\Noda}
+ 48 \pi^2 \alpha^3 C_A^2 {\Noda}^2
+ 1026 \alpha^3 C_A^2 {\Noda}^2
\right. \nonumber \\
&& \left. ~~~
+ 1296 \alpha^3 C_A {\Nda} \Nf {\Noda} T_F
- 432 \alpha^3 C_A \Nf {\Noda}^2 T_F
- 1032 \pi^2 \alpha^2 C_A^2 {\Nda}^2
\right. \nonumber \\
&& \left. ~~~
- 4374 \alpha^2 C_A^2 {\Nda}^2
+ 1252 \pi^2 \alpha^2 C_A^2 {\Nda} {\Noda}
+ 8289 \alpha^2 C_A^2 {\Nda} {\Noda}
\right. \nonumber \\
&& \left. ~~~
- 524 \pi^2 \alpha^2 C_A^2 {\Noda}^2
- 3051 \alpha^2 C_A^2 {\Noda}^2
- 512 \pi^2 \alpha^2 C_A {\Nda} \Nf {\Noda} T_F
\right. \nonumber \\
&& \left. ~~~
- 3456 \alpha^2 C_A {\Nda} \Nf {\Noda} T_F
+ 256 \pi^2 \alpha^2 C_A \Nf {\Noda}^2 T_F
+ 1728 \alpha^2 C_A \Nf {\Noda}^2 T_F
\right. \nonumber \\
&& \left. ~~~
- 432 \pi^2 \alpha C_A^2 {\Nda}^2
- 4860 \alpha C_A^2 {\Nda}^2
- 1240 \pi^2 \alpha C_A^2 {\Nda} {\Noda}
\right. \nonumber \\
&& \left. ~~~
- 1134 \alpha C_A^2 {\Nda} {\Noda}
+ 936 \pi^2 \alpha C_A^2 {\Noda}^2
+ 2106 \alpha C_A^2 {\Noda}^2
\right. \nonumber \\
&& \left. ~~~
+ 320 \pi^2 \alpha C_A {\Nda} \Nf {\Noda} T_F
+ 1296 \alpha C_A {\Nda} \Nf {\Noda} T_F
- 576 \pi^2 \alpha C_A \Nf {\Noda}^2 T_F
\right. \nonumber \\
&& \left. ~~~
- 1296 \alpha C_A \Nf {\Noda}^2 T_F
+ 720 \pi^2 C_A^2 {\Nda}^2
+ 2916 C_A^2 {\Nda}^2
\right. \nonumber \\
&& \left. ~~~
- 1296 \pi^2 C_A^2 {\Nda} {\Noda}
- 2916 C_A^2 {\Nda} {\Noda}
- 14688 C_A^2 {\Noda}^2
\right. \nonumber \\
&& \left. ~~~
+ 8640 C_A \Nf {\Noda}^2 T_F
+ 5184 C_F \Nf {\Noda}^2 T_F \right] \frac{a^3}{1296{\Noda}^2} ~+~ O(a^4) ~.
\end{eqnarray}
The anomalous dimensions are
\begin{eqnarray}
\gamma_A^{\MOMgggs}(a,\alpha) &=&
[ -~ 3 \alpha C_A {\Nda} + 3 \alpha C_A {\Noda} + 9 C_A {\Nda} - 13 C_A {\Noda} 
+ 8 \Nf {\Noda} T_F ] \frac{a}{6{\Noda}} \nonumber \\
&& + \left[ -~ 432 \psi^\prime(\third) \alpha^3 C_A^2 {\Nda}^2
+ 648 \psi^\prime(\third) \alpha^3 C_A^2 {\Nda} {\Noda}
\right. \nonumber \\
&& \left. ~~~
- 216 \psi^\prime(\third) \alpha^3 C_A^2 {\Noda}^2
+ 1836 \psi^\prime(\third) \alpha^2 C_A^2 {\Nda}^2
\right. \nonumber \\
&& \left. ~~~
- 4032 \psi^\prime(\third) \alpha^2 C_A^2 {\Nda} {\Noda}
+ 1908 \psi^\prime(\third) \alpha^2 C_A^2 {\Noda}^2
\right. \nonumber \\
&& \left. ~~~
+ 1152 \psi^\prime(\third) \alpha^2 C_A {\Nda} \Nf {\Noda} T_F
- 576 \psi^\prime(\third) \alpha^2 C_A \Nf {\Noda}^2 T_F
\right. \nonumber \\
&& \left. ~~~
- 5832 \psi^\prime(\third) \alpha C_A^2 {\Nda}^2
+ 10296 \psi^\prime(\third) \alpha C_A^2 {\Nda} {\Noda}
\right. \nonumber \\
&& \left. ~~~
- 5040 \psi^\prime(\third) \alpha C_A^2 {\Noda}^2
- 3744 \psi^\prime(\third) \alpha C_A {\Nda} \Nf {\Noda} T_F
\right. \nonumber \\
&& \left. ~~~
+ 4896 \psi^\prime(\third) \alpha C_A \Nf {\Noda}^2 T_F
+ 12636 \psi^\prime(\third) C_A^2 {\Nda}^2
\right. \nonumber \\
&& \left. ~~~
- 20736 \psi^\prime(\third) C_A^2 {\Nda} {\Noda}
+ 3588 \psi^\prime(\third) C_A^2 {\Noda}^2
\right. \nonumber \\
&& \left. ~~~
+ 18144 \psi^\prime(\third) C_A {\Nda} \Nf {\Noda} T_F
- 12192 \psi^\prime(\third) C_A \Nf {\Noda}^2 T_F
\right. \nonumber \\
&& \left. ~~~
+ 6144 \psi^\prime(\third) \Nf^2 {\Noda}^2 T_F^2
- 486 \alpha^4 C_A^2 {\Nda}^2
+ 648 \alpha^4 C_A^2 {\Nda} {\Noda}
\right. \nonumber \\
&& \left. ~~~
- 162 \alpha^4 C_A^2 {\Noda}^2
+ 288 \pi^2 \alpha^3 C_A^2 {\Nda}^2
+ 6318 \alpha^3 C_A^2 {\Nda}^2
\right. \nonumber \\
&& \left. ~~~
- 432 \pi^2 \alpha^3 C_A^2 {\Nda} {\Noda}
- 6966 \alpha^3 C_A^2 {\Nda} {\Noda}
+ 144 \pi^2 \alpha^3 C_A^2 {\Noda}^2
\right. \nonumber \\
&& \left. ~~~
+ 1674 \alpha^3 C_A^2 {\Noda}^2
+ 1296 \alpha^3 C_A {\Nda} \Nf {\Noda} T_F
- 432 \alpha^3 C_A \Nf {\Noda}^2 T_F
\right. \nonumber \\
&& \left. ~~~
- 1224 \pi^2 \alpha^2 C_A^2 {\Nda}^2
+ 9477 \alpha^2 C_A^2 {\Nda}^2
+ 2688 \pi^2 \alpha^2 C_A^2 {\Nda} {\Noda}
\right. \nonumber \\
&& \left. ~~~
+ 2025 \alpha^2 C_A^2 {\Nda} {\Noda}
- 1272 \pi^2 \alpha^2 C_A^2 {\Noda}^2
- 3078 \alpha^2 C_A^2 {\Noda}^2
\right. \nonumber \\
&& \left. ~~~
- 768 \pi^2 \alpha^2 C_A {\Nda} \Nf {\Noda} T_F
- 2592 \alpha^2 C_A {\Nda} \Nf {\Noda} T_F
\right. \nonumber \\
&& \left. ~~~
+ 384 \pi^2 \alpha^2 C_A \Nf {\Noda}^2 T_F
+ 1296 \alpha^2 C_A \Nf {\Noda}^2 T_F
+ 3888 \pi^2 \alpha C_A^2 {\Nda}^2
\right. \nonumber \\
&& \left. ~~~
+ 34020 \alpha C_A^2 {\Nda}^2
- 6864 \pi^2 \alpha C_A^2 {\Nda} {\Noda}
- 6048 \alpha C_A^2 {\Nda} {\Noda}
\right. \nonumber \\
&& \left. ~~~
+ 3360 \pi^2 \alpha C_A^2 {\Noda}^2
- 270 \alpha C_A^2 {\Noda}^2
+ 2496 \pi^2 \alpha C_A {\Nda} \Nf {\Noda} T_F
\right. \nonumber \\
&& \left. ~~~
+ 3024 \alpha C_A {\Nda} \Nf {\Noda} T_F
- 3264 \pi^2 \alpha C_A \Nf {\Noda}^2 T_F
- 3024 \alpha C_A \Nf {\Noda}^2 T_F
\right. \nonumber \\
&& \left. ~~~
- 8424 \pi^2 C_A^2 {\Nda}^2
+ 8019 C_A^2 {\Nda}^2
+ 13824 \pi^2 C_A^2 {\Nda} {\Noda}
\right. \nonumber \\
&& \left. ~~~
+ 16119 C_A^2 {\Nda} {\Noda}
- 2392 \pi^2 C_A^2 {\Noda}^2
- 5310 C_A^2 {\Noda}^2
\right. \nonumber \\
&& \left. ~~~
- 12096 \pi^2 C_A {\Nda} \Nf {\Noda} T_F
- 6480 C_A {\Nda} \Nf {\Noda} T_F
+ 8128 \pi^2 C_A \Nf {\Noda}^2 T_F
\right. \nonumber \\
&& \left. ~~~
+ 4608 C_A \Nf {\Noda}^2 T_F
+ 15552 C_F \Nf {\Noda}^2 T_F
- 4096 \pi^2 \Nf^2 {\Noda}^2 T_F^2
\right. \nonumber \\
&& \left. ~~~
+ 2304 \Nf^2 {\Noda}^2 T_F^2 \right] \frac{a^2}{3888 {\Noda}^2} ~+~ O(a^3) 
\end{eqnarray}
\begin{eqnarray}
\gamma_\alpha^{\MOMgggs}(a,\alpha) &=&
\left[ -~ 6 \alpha^2 C_A {\Nda} - 3 \alpha^2 C_A {\Noda} - 36 \alpha C_A {\Nda} 
+ 26 \alpha C_A {\Noda} - 16 \alpha \Nf {\Noda} T_F 
\right. \nonumber \\
&& \left. 
\, -~ 36 C_A {\Nda} \right] \frac{a}{12 \alpha {\Noda}} \nonumber \\
&&
+ \left[ -~ 432 \psi^\prime(\third) \alpha^4 C_A^2 {\Nda}^2
+ 108 \psi^\prime(\third) \alpha^4 C_A^2 {\Noda}^2
- 2052 \psi^\prime(\third) \alpha^3 C_A^2 {\Nda}^2
\right. \nonumber \\
&& \left. ~~~
+ 2466 \psi^\prime(\third) \alpha^3 C_A^2 {\Nda} {\Noda}
- 1422 \psi^\prime(\third) \alpha^3 C_A^2 {\Noda}^2
\right. \nonumber \\
&& \left. ~~~
- 1152 \psi^\prime(\third) \alpha^3 C_A {\Nda} \Nf {\Noda} T_F
+ 576 \psi^\prime(\third) \alpha^3 C_A \Nf {\Noda}^2 T_F
\right. \nonumber \\
&& \left. ~~~
- 3564 \psi^\prime(\third) \alpha^2 C_A^2 {\Nda}^2
- 8154 \psi^\prime(\third) \alpha^2 C_A^2 {\Nda} {\Noda}
\right. \nonumber \\
&& \left. ~~~
+ 4626 \psi^\prime(\third) \alpha^2 C_A^2 {\Noda}^2
- 864 \psi^\prime(\third) \alpha^2 C_A {\Nda} \Nf {\Noda} T_F
\right. \nonumber \\
&& \left. ~~~
- 3744 \psi^\prime(\third) \alpha^2 C_A \Nf {\Noda}^2 T_F
- 22032 \psi^\prime(\third) \alpha C_A^2 {\Nda}^2
\right. \nonumber \\
&& \left. ~~~
+ 17388 \psi^\prime(\third) \alpha C_A^2 {\Nda} {\Noda}
- 3588 \psi^\prime(\third) \alpha C_A^2 {\Noda}^2
\right. \nonumber \\
&& \left. ~~~
- 25056 \psi^\prime(\third) \alpha C_A {\Nda} \Nf {\Noda} T_F
+ 12192 \psi^\prime(\third) \alpha C_A \Nf {\Noda}^2 T_F
\right. \nonumber \\
&& \left. ~~~
- 6144 \psi^\prime(\third) \alpha \Nf^2 {\Noda}^2 T_F^2
- 25272 \psi^\prime(\third) C_A^2 {\Nda}^2
\right. \nonumber \\
&& \left. ~~~
+ 4968 \psi^\prime(\third) C_A^2 {\Nda} {\Noda}
- 13824 \psi^\prime(\third) C_A {\Nda} \Nf {\Noda} T_F
- 486 \alpha^5 C_A^2 {\Nda}^2
\right. \nonumber \\
&& \left. ~~~
- 81 \alpha^5 C_A^2 {\Nda} {\Noda}
+ 81 \alpha^5 C_A^2 {\Noda}^2
+ 288 \pi^2 \alpha^4 C_A^2 {\Nda}^2
- 1944 \alpha^4 C_A^2 {\Nda}^2
\right. \nonumber \\
&& \left. ~~~
+ 3078 \alpha^4 C_A^2 {\Nda} {\Noda}
- 72 \pi^2 \alpha^4 C_A^2 {\Noda}^2
- 945 \alpha^4 C_A^2 {\Noda}^2
\right. \nonumber \\
&& \left. ~~~
- 1296 \alpha^4 C_A {\Nda} \Nf {\Noda} T_F
+ 432 \alpha^4 C_A \Nf {\Noda}^2 T_F
+ 1368 \pi^2 \alpha^3 C_A^2 {\Nda}^2
\right. \nonumber \\
&& \left. ~~~
- 6804 \alpha^3 C_A^2 {\Nda}^2
- 1644 \pi^2 \alpha^3 C_A^2 {\Nda} {\Noda}
- 7614 \alpha^3 C_A^2 {\Nda} {\Noda}
\right. \nonumber \\
&& \left. ~~~
+ 948 \pi^2 \alpha^3 C_A^2 {\Noda}^2
+ 4050 \alpha^3 C_A^2 {\Noda}^2
+ 768 \pi^2 \alpha^3 C_A {\Nda} \Nf {\Noda} T_F
\right. \nonumber \\
&& \left. ~~~
+ 2592 \alpha^3 C_A {\Nda} \Nf {\Noda} T_F
- 384 \pi^2 \alpha^3 C_A \Nf {\Noda}^2 T_F
\right. \nonumber \\
&& \left. ~~~
- 1296 \alpha^3 C_A \Nf {\Noda}^2 T_F
+ 2376 \pi^2 \alpha^2 C_A^2 {\Nda}^2
- 49086 \alpha^2 C_A^2 {\Nda}^2
\right. \nonumber \\
&& \left. ~~~
+ 5436 \pi^2 \alpha^2 C_A^2 {\Nda} {\Noda}
+ 8775 \alpha^2 C_A^2 {\Nda} {\Noda}
- 3084 \pi^2 \alpha^2 C_A^2 {\Noda}^2
\right. \nonumber \\
&& \left. ~~~
- 108 \alpha^2 C_A^2 {\Noda}^2
+ 576 \pi^2 \alpha^2 C_A {\Nda} \Nf {\Noda} T_F
- 4752 \alpha^2 C_A {\Nda} \Nf {\Noda} T_F
\right. \nonumber \\
&& \left. ~~~
+ 2496 \pi^2 \alpha^2 C_A \Nf {\Noda}^2 T_F
+ 3456 \alpha^2 C_A \Nf {\Noda}^2 T_F
+ 14688 \pi^2 \alpha C_A^2 {\Nda}^2
\right. \nonumber \\
&& \left. ~~~
- 10206 \alpha C_A^2 {\Nda}^2
- 11592 \pi^2 \alpha C_A^2 {\Nda} {\Noda}
- 22599 \alpha C_A^2 {\Nda} {\Noda}
\right. \nonumber \\
&& \left. ~~~
+ 2392 \pi^2 \alpha C_A^2 {\Noda}^2
+ 5310 \alpha C_A^2 {\Noda}^2
+ 16704 \pi^2 \alpha C_A {\Nda} \Nf {\Noda} T_F
\right. \nonumber \\
&& \left. ~~~
+ 15552 \alpha C_A {\Nda} \Nf {\Noda} T_F
- 8128 \pi^2 \alpha C_A \Nf {\Noda}^2 T_F
\right. \nonumber \\
&& \left. ~~~
- 4608 \alpha C_A \Nf {\Noda}^2 T_F
- 15552 \alpha C_F \Nf {\Noda}^2 T_F
+ 4096 \pi^2 \alpha \Nf^2 {\Noda}^2 T_F^2
\right. \nonumber \\
&& \left. ~~~
- 2304 \alpha \Nf^2 {\Noda}^2 T_F^2
+ 16848 \pi^2 C_A^2 {\Nda}^2
+ 116640 C_A^2 {\Nda}^2
\right. \nonumber \\
&& \left. ~~~
- 3312 \pi^2 C_A^2 {\Nda} {\Noda}
- 14904 C_A^2 {\Nda} {\Noda}
+ 9216 \pi^2 C_A {\Nda} \Nf {\Noda} T_F
\right. \nonumber \\
&& \left. ~~~
+ 10368 C_A {\Nda} \Nf {\Noda} T_F \right] 
\frac{a^2}{3888 \alpha {\Noda}^2} ~+~ O(a^3) 
\end{eqnarray}
\begin{eqnarray}
\gamma_c^{\MOMgggs}(a,\alpha) &=&
\left[ -~ 2 \alpha {\Nda} + \alpha {\Noda} - 6 {\Nda} - 3 {\Noda} \right]
\frac{C_Aa}{4 {\Noda}} \nonumber \\
&& + \left[ -~ 288 \psi^\prime(\third) \alpha^3 C_A {\Nda}^2
+ 288 \psi^\prime(\third) \alpha^3 C_A {\Nda} {\Noda}
- 72 \psi^\prime(\third) \alpha^3 C_A {\Noda}^2
\right. \nonumber \\
&& \left. ~~~
- 504 \psi^\prime(\third) \alpha^2 C_A {\Nda}^2
- 828 \psi^\prime(\third) \alpha^2 C_A {\Nda} {\Noda}
+ 540 \psi^\prime(\third) \alpha^2 C_A {\Noda}^2
\right. \nonumber \\
&& \left. ~~~
- 1728 \psi^\prime(\third) \alpha C_A {\Nda}^2
+ 552 \psi^\prime(\third) \alpha C_A {\Nda} {\Noda}
- 1248 \psi^\prime(\third) \alpha C_A {\Noda}^2
\right. \nonumber \\
&& \left. ~~~
- 1536 \psi^\prime(\third) \alpha {\Nda} \Nf {\Noda} T_F
+ 768 \psi^\prime(\third) \alpha \Nf {\Noda}^2 T_F
- 8424 \psi^\prime(\third) C_A {\Nda}^2
\right. \nonumber \\
&& \left. ~~~
- 2556 \psi^\prime(\third) C_A {\Nda} {\Noda}
+ 828 \psi^\prime(\third) C_A {\Noda}^2
- 4608 \psi^\prime(\third) {\Nda} \Nf {\Noda} T_F
\right. \nonumber \\
&& \left. ~~~
- 2304 \psi^\prime(\third) \Nf {\Noda}^2 T_F
- 324 \alpha^4 C_A {\Nda}^2
+ 270 \alpha^4 C_A {\Nda} {\Noda}
\right. \nonumber \\
&& \left. ~~~
- 54 \alpha^4 C_A {\Noda}^2
+ 192 \pi^2 \alpha^3 C_A {\Nda}^2
+ 972 \alpha^3 C_A {\Nda}^2
\right. \nonumber \\
&& \left. ~~~
- 192 \pi^2 \alpha^3 C_A {\Nda} {\Noda}
- 2106 \alpha^3 C_A {\Nda} {\Noda}
+ 48 \pi^2 \alpha^3 C_A {\Noda}^2
\right. \nonumber \\
&& \left. ~~~
+ 648 \alpha^3 C_A {\Noda}^2
+ 336 \pi^2 \alpha^2 C_A {\Nda}^2
+ 5184 \alpha^2 C_A {\Nda}^2
\right. \nonumber \\
&& \left. ~~~
+ 552 \pi^2 \alpha^2 C_A {\Nda} {\Noda}
- 1944 \alpha^2 C_A {\Nda} {\Noda}
- 360 \pi^2 \alpha^2 C_A {\Noda}^2
\right. \nonumber \\
&& \left. ~~~
- 648 \alpha^2 C_A {\Noda}^2
+ 1152 \pi^2 \alpha C_A {\Nda}^2
- 10368 \alpha C_A {\Nda}^2
\right. \nonumber \\
&& \left. ~~~
- 368 \pi^2 \alpha C_A {\Nda} {\Noda}
- 144 \alpha C_A {\Nda} {\Noda}
+ 832 \pi^2 \alpha C_A {\Noda}^2
\right. \nonumber \\
&& \left. ~~~
- 1710 \alpha C_A {\Noda}^2
+ 1024 \pi^2 \alpha {\Nda} \Nf {\Noda} T_F
- 576 \alpha {\Nda} \Nf {\Noda} T_F
\right. \nonumber \\
&& \left. ~~~
- 512 \pi^2 \alpha \Nf {\Noda}^2 T_F
+ 288 \alpha \Nf {\Noda}^2 T_F
+ 5616 \pi^2 C_A {\Nda}^2
+ 17010 C_A {\Nda}^2
\right. \nonumber \\
&& \left. ~~~
+ 1704 \pi^2 C_A {\Nda} {\Noda}
- 1485 C_A {\Nda} {\Noda}
- 552 \pi^2 C_A {\Noda}^2
- 378 C_A {\Noda}^2
\right. \nonumber \\
&& \left. ~~~
+ 3072 \pi^2 {\Nda} \Nf {\Noda} T_F
+ 864 {\Nda} \Nf {\Noda} T_F
+ 1536 \pi^2 \Nf {\Noda}^2 T_F
\right. \nonumber \\
&& \left. ~~~
+ 432 \Nf {\Noda}^2 T_F \right] \frac{C_A a^2}{2592 {\Noda}^2} ~+~ O(a^3) 
\end{eqnarray}
and
\begin{eqnarray}
\gamma_\psi^{\MOMgggs}(a,\alpha) &=&
\frac{ \alpha {\Noda} T_F a}{\NF} \nonumber \\
&& + \left[ 72 \psi^\prime(\third) \alpha^3 C_A C_F \NF
- 108 \psi^\prime(\third) \alpha^3 C_A {\Noda} T_F
- 90 \psi^\prime(\third) \alpha^2 C_A C_F \NF
\right. \nonumber \\
&& \left. ~~~
+ 252 \psi^\prime(\third) \alpha^2 C_A {\Noda} T_F
+ 702 \psi^\prime(\third) \alpha C_A C_F \NF
- 840 \psi^\prime(\third) \alpha C_A {\Noda} T_F
\right. \nonumber \\
&& \left. ~~~
+ 384 \psi^\prime(\third) \alpha \Nf {\Noda} T_F^2
+ 81 \alpha^4 C_A C_F \NF
- 108 \alpha^4 C_A {\Noda} T_F
\right. \nonumber \\
&& \left. ~~~
- 48 \pi^2 \alpha^3 C_A C_F \NF 
- 486 \alpha^3 C_A C_F \NF
+ 72 \pi^2 \alpha^3 C_A {\Noda} T_F
\right. \nonumber \\
&& \left. ~~~
+ 729 \alpha^3 C_A {\Noda} T_F
+ 60 \pi^2 \alpha^2 C_A C_F \NF
- 162 \alpha^2 C_A C_F \NF
\right. \nonumber \\
&& \left. ~~~
- 168 \pi^2 \alpha^2 C_A {\Noda} T_F
+ 324 \alpha^2 C_A {\Noda} T_F
- 468 \pi^2 \alpha C_A C_F \NF
\right. \nonumber \\
&& \left. ~~~
+ 648 \alpha C_A C_F \NF
+ 560 \pi^2 \alpha C_A {\Noda} T_F
- 1017 \alpha C_A {\Noda} T_F
\right. \nonumber \\
&& \left. ~~~
- 256 \pi^2 \alpha \Nf {\Noda} T_F^2
+ 144 \alpha \Nf {\Noda} T_F^2
+ 2025 C_A C_F \NF
\right. \nonumber \\
&& \left. ~~~
- 486 C_F^2 \NF
- 648 C_F \NF \Nf T_F \right] \frac{a^2}{324 \NF} ~+~ O(a^3) ~. 
\end{eqnarray}

The explicit forms of the associated amplitudes are
\begin{eqnarray}
\left. \Sigma^{\mbox{\footnotesize{ggg}}}_{(1)}(p,q) \right|_{\MOMgggs} &=&
\left. \Sigma^{\mbox{\footnotesize{ggg}}}_{(2)}(p,q) \right|_{\MOMgggs} ~=~
-~ \frac{1}{2} \left. \Sigma^{\mbox{\footnotesize{ggg}}}_{(3)}(p,q)
\right|_{\MOMgggs} \nonumber \\
&=& -~ \left. \Sigma^{\mbox{\footnotesize{ggg}}}_{(4)}(p,q) \right|_{\MOMgggs}
~=~ \frac{1}{2} \left. \Sigma^{\mbox{\footnotesize{ggg}}}_{(5)}(p,q)
\right|_{\MOMgggs} \nonumber \\
&=& -~ \left. \Sigma^{\mbox{\footnotesize{ggg}}}_{(6)}(p,q) \right|_{\MOMgggs}
~=~ 1 + O(a^2) \nonumber \\
\left. \Sigma^{\mbox{\footnotesize{ggg}}}_{(7)}(p,q) \right|_{\MOMgggs} &=&
2 \left. \Sigma^{\mbox{\footnotesize{ggg}}}_{(9)}(p,q) \right|_{\MOMgggs} ~=~
-~ 2 \left. \Sigma^{\mbox{\footnotesize{ggg}}}_{(11)}(p,q)
\right|_{\MOMgggs} \nonumber \\
&=& -~ \left. \Sigma^{\mbox{\footnotesize{ggg}}}_{(14)}(p,q) \right|_{\MOMgggs}
\nonumber \\
&=& \left[ 
108 \psi^\prime(\third) \alpha^5 C_A {\Nda}
- 36 \psi^\prime(\third) \alpha^5 C_A {\Noda}
- 324 \psi^\prime(\third) \alpha^4 C_A {\Nda}
\right. \nonumber \\
&& \left. \,
+~ 162 \psi^\prime(\third) \alpha^4 C_A {\Noda}
+ 324 \psi^\prime(\third) \alpha^3 C_A {\Nda}
- 108 \psi^\prime(\third) \alpha^3 C_A {\Noda}
\right. \nonumber \\
&& \left. \,
+~ 1296 \psi^\prime(\third) \alpha^2 C_A {\Nda}
- 456 \psi^\prime(\third) \alpha^2 C_A {\Noda}
+ 768 \psi^\prime(\third) \alpha^2 \Nf {\Noda} T_F
\right. \nonumber \\
&& \left. \,
+~ 216 \psi^\prime(\third) \alpha C_A {\Nda}
+ 270 \psi^\prime(\third) C_A {\Nda}
- 72 \pi^2 \alpha^5 C_A {\Nda}
- 324 \alpha^5 C_A {\Nda}
\right. \nonumber \\
&& \left. \,
+~ 24 \pi^2 \alpha^5 C_A {\Noda}
+ 108 \alpha^5 C_A {\Noda}
+ 216 \pi^2 \alpha^4 C_A {\Nda}
+ 810 \alpha^4 C_A {\Nda}
\right. \nonumber \\
&& \left. \,
-~ 108 \pi^2 \alpha^4 C_A {\Noda}
- 405 \alpha^4 C_A {\Noda}
- 216 \pi^2 \alpha^3 C_A {\Nda}
- 1377 \alpha^3 C_A {\Nda}
\right. \nonumber \\
&& \left. \,
+~ 72 \pi^2 \alpha^3 C_A {\Noda}
+ 1458 \alpha^3 C_A {\Noda}
- 864 \pi^2 \alpha^2 C_A {\Nda}
+ 891 \alpha^2 C_A {\Nda}
\right. \nonumber \\
&& \left. \,
+~ 304 \pi^2 \alpha^2 C_A {\Noda}
- 873 \alpha^2 C_A {\Noda}
- 512 \pi^2 \alpha^2 \Nf {\Noda} T_F
- 576 \alpha^2 \Nf {\Noda} T_F
\right. \nonumber \\
&& \left. \,
-~ 144 \pi^2 \alpha C_A {\Nda}
- 243 \alpha C_A {\Nda}
- 180 \pi^2 C_A {\Nda}
+ 243 C_A {\Nda} \right] \frac{a}{972 \alpha^2 {\Noda}} \nonumber \\
&& +~ O(a^2) 
\nonumber \\
\left. \Sigma^{\mbox{\footnotesize{ggg}}}_{(8)}(p,q) \right|_{\MOMgggs} &=&
-~ \left. \Sigma^{\mbox{\footnotesize{ggg}}}_{(13)}(p,q) \right|_{\MOMgggs}
\nonumber \\
&=& \left[ 
-~ 108 \psi^\prime(\third) \alpha^5 C_A {\Nda}
+ 36 \psi^\prime(\third) \alpha^5 C_A {\Noda}
+ 540 \psi^\prime(\third) \alpha^4 C_A {\Nda}
\right. \nonumber \\
&& \left. \,
-~ 270 \psi^\prime(\third) \alpha^4 C_A {\Noda}
- 270 \psi^\prime(\third) \alpha^3 C_A {\Nda}
+ 378 \psi^\prime(\third) \alpha^3 C_A {\Noda}
\right. \nonumber \\
&& \left. \,
+~ 1242 \psi^\prime(\third) \alpha^2 C_A {\Nda}
- 390 \psi^\prime(\third) \alpha^2 C_A {\Noda}
+ 384 \psi^\prime(\third) \alpha^2 \Nf {\Noda} T_F
\right. \nonumber \\
&& \left. \,
+~ 216 \psi^\prime(\third) \alpha C_A {\Nda}
+ 270 \psi^\prime(\third) C_A {\Nda}
+ 72 \pi^2 \alpha^5 C_A {\Nda}
\right. \nonumber \\
&& \left. \,
+~ 567 \alpha^5 C_A {\Nda}
- 24 \pi^2 \alpha^5 C_A {\Noda}
- 189 \alpha^5 C_A {\Noda}
- 360 \pi^2 \alpha^4 C_A {\Nda}
\right. \nonumber \\
&& \left. \,
-~ 2268 \alpha^4 C_A {\Nda}
+ 180 \pi^2 \alpha^4 C_A {\Noda}
+ 1134 \alpha^4 C_A {\Noda}
+ 180 \pi^2 \alpha^3 C_A {\Nda}
\right. \nonumber \\
&& \left. \,
+~ 648 \alpha^3 C_A {\Nda}
- 252 \pi^2 \alpha^3 C_A {\Noda}
- 243 \alpha^3 C_A {\Noda}
\right. \nonumber \\
&& \left. \,
-~ 828 \pi^2 \alpha^2 C_A {\Nda}
+ 1053 \alpha^2 C_A {\Nda}
+ 260 \pi^2 \alpha^2 C_A {\Noda}
- 1206 \alpha^2 C_A {\Noda}
\right. \nonumber \\
&& \left. \,
-~ 256 \pi^2 \alpha^2 \Nf {\Noda} T_F
+ 1008 \alpha^2 \Nf {\Noda} T_F
- 144 \pi^2 \alpha C_A {\Nda}
- 243 \alpha C_A {\Nda}
\right. \nonumber \\
&& \left. \,
-~ 180 \pi^2 C_A {\Nda}
+ 243 C_A {\Nda} \right] \frac{a}{972 \alpha^2 {\Noda}} ~+~ O(a^2)
\nonumber \\
\left. \Sigma^{\mbox{\footnotesize{ggg}}}_{(10)}(p,q) \right|_{\MOMgggs} &=&
-~ \left. \Sigma^{\mbox{\footnotesize{ggg}}}_{(12)}(p,q) \right|_{\MOMgggs}
\nonumber \\
&=& \left[ 
216 \psi^\prime(\third) \alpha^3 C_A {\Nda}
- 72 \psi^\prime(\third) \alpha^3 C_A {\Noda}
- 864 \psi^\prime(\third) \alpha^2 C_A {\Nda}
\right. \nonumber \\
&& \left. \,
+~ 432 \psi^\prime(\third) \alpha^2 C_A {\Noda}
+ 594 \psi^\prime(\third) \alpha C_A {\Nda}
- 486 \psi^\prime(\third) \alpha C_A {\Noda}
\right. \nonumber \\
&& \left. \,
+~ 54 \psi^\prime(\third) C_A {\Nda}
- 66 \psi^\prime(\third) C_A {\Noda}
+ 384 \psi^\prime(\third) \Nf {\Noda} T_F
\right. \nonumber \\
&& \left. \,
-~ 144 \pi^2 \alpha^3 C_A {\Nda}
- 891 \alpha^3 C_A {\Nda}
+ 48 \pi^2 \alpha^3 C_A {\Noda}
+ 297 \alpha^3 C_A {\Noda}
\right. \nonumber \\
&& \left. \,
+~ 576 \pi^2 \alpha^2 C_A {\Nda}
+ 3078 \alpha^2 C_A {\Nda}
- 288 \pi^2 \alpha^2 C_A {\Noda}
- 1539 \alpha^2 C_A {\Noda}
\right. \nonumber \\
&& \left. \,
-~ 396 \pi^2 \alpha C_A {\Nda}
- 2025 \alpha C_A {\Nda}
+ 324 \pi^2 \alpha C_A {\Noda}
+ 1701 \alpha C_A {\Noda}
\right. \nonumber \\
&& \left. \,
-~ 36 \pi^2 C_A {\Nda}
- 162 C_A {\Nda}
+ 44 \pi^2 C_A {\Noda}
+ 333 C_A {\Noda}
\right. \nonumber \\
&& \left. \,
-~ 256 \pi^2 \Nf {\Noda} T_F
- 1584 \Nf {\Noda} T_F \right] \frac{a}{972 {\Noda}} ~+~ O(a^2) ~. 
\end{eqnarray}
Again we observe that the same symmetries emerge as in the $\MSbar$ case which
is a minor check on the computation. 

\sect{$\MOMh$ scheme.}

Having recorded the results for the triple gluon vertex at length we briefly
present the results for the ghost-gluon vertex in numerical form in order to
save space. The full analytic expressions are given in the attached data file.
Given the nature of the $\MOMh$ scheme the amplitudes are effectively trivial
since 
\begin{equation}
\left. \Sigma^{\mbox{\footnotesize{ccg}}}_{(1)}(p,q) \right|_{\MOMhs} ~=~
-~ \left. \Sigma^{\mbox{\footnotesize{ccg}}}_{(2)}(p,q) \right|_{\MOMhs} ~=~ 
\frac{1}{2} ~+~ O(a^2) ~.
\end{equation}
This is because of the symmetry of the original ghost-gluon vertex and the
definition of the $\MOMh$ scheme. With the coupling constant conversion 
function for $SU(3)$  
\begin{equation} 
C^{\MOMhs}_g(a,\alpha) ~=~ 
1 + \left[ - 0.125000 \alpha^2 - \alpha + 0.555556 \Nf - 10.432318 
\right] a ~+~ O(a^2)
\end{equation}
we can deduce the two loop $\MOMh$ anomalous dimensions. They are 
\begin{eqnarray}
\beta^{\MOMhs}(a,\alpha) &=& \left[ 0.666667 \Nf - 11.000000 \right] a^2
\nonumber \\
&& + \left[ -~ 0.625000 \alpha^3 - 0.333333 \alpha^2 \Nf - 0.750000 \alpha^2 
- 1.333333 \alpha \Nf \right. \nonumber \\
&& \left. ~~~+~ 5.500000 \alpha + 12.666667 \Nf - 108.000000 
\right] a^3 ~+~ O(a^4) \nonumber \\
\gamma_A^{\MOMhs}(a,\alpha) &=& \left[ \alpha + 0.666667 \Nf - 5.000000 
\right] a \nonumber \\
&& + \left[ -~ 0.625000 \alpha^3 - 0.333333 \alpha^2 \Nf + 0.125000 \alpha^2 
- 1.333333 \alpha \Nf \right. \nonumber \\
&& \left. ~~~+~ 7.093698 \alpha + 5.479132 \Nf - 3.468488 
\right] a^2 ~+~ O(a^3) \nonumber \\
\gamma_\alpha^{\MOMhs}(a,\alpha) &=&
\left[ - 1.250000 \alpha^2 - 0.666667 \alpha \Nf + 3.500000 \alpha - 3.000000
\right] \frac{a}{\alpha} \nonumber \\
&& + \left[ 0.625000 \alpha^4 + 0.333333 \alpha^3 \Nf - 0.250000 \alpha^3 
+ 1.333333 \alpha^2 \Nf - 7.023372 \alpha^2 \right. \nonumber \\
&& \left. ~~~-~ 3.979132 \alpha \Nf - 2.297058 \alpha + 2.000000 \Nf 
+ 15.093907 \right] \frac{a^2}{\alpha} ~+~ O(a^3) \nonumber \\
\gamma_c^{\MOMhs}(a,\alpha) &=& \left[ 0.250000 \alpha - 3.750000 \right] a
\nonumber \\
&& + \left[ 0.187500 \alpha^2 - 7.945326 \alpha + 1.250000 \Nf 
- 0.726366 \right] a^2 ~+~ O(a^3) \nonumber \\
\gamma_\psi^{\MOMhs}(a,\alpha) &=& \alpha a 
+ \left[ 0.500000 \alpha^2 - 0.281302 \alpha - 1.333333 \Nf + 22.333333 
\right] a^2 \nonumber \\
&& +~ O(a^3)
\end{eqnarray}
which agree with the explicit direct two loop computation. 

\sect{$\MOMq$ scheme.}

For the $\MOMq$ scheme we also give the results in numerical form for $SU(3)$. 
Though the non-channel $1$ amplitudes are non-trivial here since
\begin{eqnarray}
\left. \Sigma^{\mbox{\footnotesize{qqg}}}_{(1)}(p,q) \right|_{\MOMqs} &=&
-~ 1.000000 + O(a^2) \nonumber \\
\left. \Sigma^{\mbox{\footnotesize{qqg}}}_{(2)}(p,q) \right|_{\MOMqs} &=&
\left. \Sigma^{\mbox{\footnotesize{qqg}}}_{(5)}(p,q) \right|_{\MOMqs} ~=~
\left[ 0.138008 \alpha^2 + 1.562605 \alpha - 1.540716 \right] a ~+~ O(a^2)
\nonumber \\
\left. \Sigma^{\mbox{\footnotesize{qqg}}}_{(3)}(p,q) \right|_{\MOMqs} &=&
\left. \Sigma^{\mbox{\footnotesize{qqg}}}_{(4)}(p,q) \right|_{\MOMqs} ~=~
\left[ 0.166667 \alpha^2 + 1.781302 \alpha - 0.826284 \right] a ~+~ O(a^2)
\nonumber \\
\left. \Sigma^{\mbox{\footnotesize{qqg}}}_{(6)}(p,q) \right|_{\MOMqs} &=&
\left[ 0.195326 \alpha^2 + 1.562605 \alpha + 3.971621 \right] a ~+~ O(a^2) ~.
\end{eqnarray}
The associated coupling constant conversion function is
\begin{equation} 
C^{\MOMqs}_g(a,\alpha) ~=~ 1 + \left[ 0.027337 \alpha^2 + 0.843907 \alpha 
+ 0.555556 \Nf - 7.381259 \right]  a ~+~ O(a^2)
\end{equation} 
from which we deduce that the two loop $SU(3)$ renormalization group functions 
are
\begin{eqnarray}
\beta^{\MOMqs}(a,\alpha) &=& \left[ 0.666667 \Nf - 11.000000 \right] a^2 
\nonumber \\
&& + \left[ 0.136686 \alpha^3 + 0.072899 \alpha^2 \Nf + 1.727047 \alpha^2 
+ 1.125210 \alpha \Nf \right. \nonumber \\
&& \left. ~~~-~ 5.579304 \alpha + 12.666667 \Nf - 96.936557 
\right] a^3 ~+~ O(a^4) \nonumber \\
\gamma_A^{\MOMqs}(a,\alpha) &=& \left[ \alpha + 0.666667 \Nf 
- 5.000000 \right] a \nonumber \\
&& + \left[ -~ 0.320326 \alpha^3 - 0.130217 \alpha^2 \Nf + 2.289442 \alpha^2 
+ 1.125210 \alpha \Nf \right. \nonumber \\
&& \left. ~~~-~ 5.243258 \alpha + 9.547210 \Nf - 33.979074 
\right] a^2 ~+~ O(a^3) \nonumber \\
\gamma_\alpha^{\MOMqs}(a,\alpha) &=& 
\left[ - 1.250000 \alpha^2 - 0.666667 \alpha \Nf + 3.500000 \alpha - 3.000000
\right] \frac{a}{\alpha} \nonumber \\ 
&& + \left[ 0.244157 \alpha^4 + 0.130217 \alpha^3 \Nf - 3.793408 \alpha^3 
- 1.125210 \alpha^2 \Nf - 2.657691 \alpha^2 \right. \nonumber \\
&& \left. ~~~-~ 8.047210 \alpha \Nf 
+ 7.996908 \alpha + 2.000000 \Nf - 3.212444 \right] \frac{a^2}{\alpha} ~+~
O(a^3) \nonumber \\
\gamma_c^{\MOMqs}(a,\alpha) &=& \left[ 0.250000 \alpha - 3.750000 \right] a
\nonumber \\
&& + \left[ 0.076169 \alpha^3 - 0.033075 \alpha^2 - 20.249101 \alpha 
+ 1.250000 \Nf \right. \nonumber \\
&& \left. ~~~-~ 23.609305 \right] a^2 ~+~ O(a^3) \nonumber \\
\gamma_\psi^{\MOMqs}(a,\alpha) &=& \alpha a + \left[ 0.304674 \alpha^3 
+ 4.187814 \alpha^2 + 5.820815 \alpha - 1.333333 \Nf + 22.333333 
\right] a^2 \nonumber \\
&& +~ O(a^3) ~.
\end{eqnarray} 
Unlike $\MOMh$ the quark anomalous dimension is cubic in the gauge parameter.

\sect{Curci-Ferrari gauge.} 

One interesting property of the maximal abelian gauge is that in the formal
limit $\Nda/\Noda$~$\rightarrow$~$0$ the Lagrangian becomes equivalent to gauge
fixing QCD in the nonlinear Curci-Ferrari gauge, \cite{31}. More specifically
the Lagrangian for the choice of the Curci-Ferrari gauge is, \cite{31}, 
\begin{eqnarray}
L^{\mbox{\footnotesize{CF}}} &=& -~ \frac{1}{4} G_{\mu\nu}^a
G^{a \, \mu\nu} ~-~ \frac{1}{2\alpha} (\partial^\mu A^a_\mu)^2 ~-~ 
\bar{c}^a \partial^\mu D_\mu c^a ~+~ i \bar{\psi}^{iI} \Dslash \psi^{iI} 
\nonumber \\
&& +~ \frac{g}{2} f^{abc} \partial^\mu A^a_\mu \, \bar{c}^b c^c ~+~
\frac{\alpha g^2}{8} f^{eab} f^{ecd} \bar{c}^a c^b \bar{c}^c c^d ~. 
\label{lagcf}
\end{eqnarray}
Here the colour indices have the formal range $1$~$\leq$~$a$~$\leq$~$\NA$ where
$\NA$ is the dimension of the adjoint representation of the colour group and 
$\alpha$ is the associated gauge parameter. This gauge choice differs from the 
usual linear covariant gauge fixed Lagrangian in that there is a quartic ghost 
vertex and the ghost-gluon vertex is different. The former vertex does not 
exclude renormalizability which can be seen using simple power counting with 
the proof given in \cite{20,22,23,24,26,29} which provides the relations 
between the various renormalization constants. The gauge parameter differs also
from that of the linear gauge fixing but when $\alpha$~$=$~$0$ then 
(\ref{lagcf}) reduces to the Landau gauge Lagrangian, \cite{31}. It is 
straightforward to deduce that the Curci-Ferrari gauge is a particular limit of
(\ref{lagmag}) by examining that Lagrangian and omitting any interaction with a
centre field. In some respects given the non-renormalization of certain aspects
of those fields the centre could be regarded as analogous to a background field
in the context of the background field gauge, \cite{56,57,58,59,60,61,62,63}.

Given the close relation between this gauge and the MAG, we have renormalized 
(\ref{lagcf}) in the three MOM schemes defined by the $3$-point vertices. We 
did this directly and independently of the MAG and its limit to the 
Curci-Ferrari gauge in order to have an independent check on our computations. 
In other words we verify that the limit of the MAG renormalization group 
functions and amplitudes agree when $\Nda/\Noda$~$\rightarrow$~$0$. Therefore, 
for completeness we record the direct results for the Curci-Ferrari gauge in 
the three MOM schemes. First, we recall that the relevant two loop $\MSbar$ 
renormalization group functions are, \cite{64,65},  
\begin{eqnarray}
\gamma_A(a) &=& \left[ ( 3\alpha - 13 ) C_A + 8T_F \Nf \right] \frac{a}{6}
\nonumber \\
&& +~ \left[ \left( \alpha^2 + 11\alpha - 59 \right) C_A^2 + 40 C_A T_F \Nf
+ 32 C_F T_F \Nf \right] \frac{a^2}{8} ~+~ O(a^3) \nonumber \\
\gamma_\alpha(a) &=& -~ \left[ ( 3\alpha - 26 ) C_A + 16 T_F \Nf \right]
\frac{a}{12} \nonumber \\
&& -~ \left[ \left( \alpha^2 + 17\alpha - 118 \right) C_A^2 + 80 C_A T_F \Nf
+ 64 C_F T_F \Nf \right] \frac{a^2}{16} ~+~ O(a^3) \nonumber \\
\gamma_c(a) &=& ( \alpha - 3 ) C_A \frac{a}{4} ~+~ \left[ \left( 3\alpha^2
- 3\alpha - 95 \right) C_A^2 + 40 C_A T_F \Nf \right] \frac{a^2}{48} ~+~ O(a^3)
\nonumber \\
\gamma_\psi(a) &=& \frac{\alpha C_F}{4} a
+ \frac{1}{4} \left[ ( 8 \alpha + 25 ) C_A C_F - 6 C_F^2 - 8 C_F T_F \Nf
\right] a^2 ~+~ O(a^3) ~. 
\end{eqnarray}
For the MOM results we first record the mappings for the parameters at one
loop for each scheme. The three coupling constant mappings are 
\begin{eqnarray}
a_{\MOMgggs} &=&
a + \left[ 36 \psi^\prime(\third) \alpha^2 C_A
- 162 \psi^\prime(\third) \alpha C_A
+ 138 \psi^\prime(\third) C_A
- 384 \psi^\prime(\third) \Nf T_F
\right. \nonumber \\
&& \left. ~~~~~\,
+ 27 \alpha^3 C_A
- 24 \pi^2 \alpha^2 C_A 
- 162 \alpha^2 C_A
+ 108 \pi^2 \alpha C_A 
+ 243 \alpha C_A
\right. \nonumber \\
&& \left. ~~~~~\,
- 92 \pi^2 C_A 
+ 2376 C_A
+ 256 \pi^2 \Nf T_F
- 864 \Nf T_F \right] \frac{a^2}{324} ~+~ O(a^3) \nonumber \\
a_{\MOMhs} &=&
a + \left[ - \, 12 \psi^\prime(\third) \alpha C_A
+ 30 \psi^\prime(\third) C_A
+ 27 \alpha^2 C_A
+ 8 \pi^2 \alpha C_A 
\right. \nonumber \\
&& \left. ~~~~~\,
+ 108 \alpha C_A
- 20 \pi^2 C_A 
+ 669 C_A
- 240 \Nf T_F \right] \frac{a^2}{108} ~+~ O(a^3) \nonumber \\
a_{\MOMqs} &=&
a + \left[ 6 \psi^\prime(\third) \alpha^2 C_A
- 24 \psi^\prime(\third) \alpha C_A
- 96 \psi^\prime(\third) \alpha C_F
- 78 \psi^\prime(\third) C_A
\right. \nonumber \\
&& \left. ~~~~~\,
+ 48 \psi^\prime(\third) C_F
- 4 \pi^2 \alpha^2 C_A 
- 27 \alpha^2 C_A
+ 16 \pi^2 \alpha C_A 
+ 54 \alpha C_A
\right. \nonumber \\
&& \left. ~~~~~\,
+ 64 \pi^2 \alpha C_F 
+ 216 \alpha C_F
+ 52 \pi^2 C_A 
+ 993 C_A
- 32 \pi^2 C_F 
- 432 C_F
\right. \nonumber \\
&& \left. ~~~~~\,
- 240 \Nf T_F \right] \frac{a^2}{108} ~+~ O(a^3) ~.
\end{eqnarray}
Subsequently we can deduce that the coupling constant conversion functions in 
each scheme are 
\begin{eqnarray}
C_g^{\MOMgggs}(a,\alpha) &=& 
1 + \left[ - \, 36 \psi^\prime(\third) \alpha^2 C_A
+ 162 \psi^\prime(\third) \alpha C_A
- 138 \psi^\prime(\third) C_A
+ 384 \psi^\prime(\third) \Nf T_F
\right. \nonumber \\
&& \left. ~~~~~\,
- 27 \alpha^3 C_A
+ 24 \pi^2 \alpha^2 C_A 
+ 162 \alpha^2 C_A
- 108 \pi^2 \alpha C_A 
- 243 \alpha C_A
\right. \nonumber \\
&& \left. ~~~~~\,
+ 92 \pi^2 C_A 
- 2376 C_A
- 256 \pi^2 \Nf T_F
+ 864 \Nf T_F \right] \frac{a}{648} ~+~ O(a^2) \nonumber \\
C_g^{\MOMhs}(a,\alpha) &=& 
1 + \left[ 12 \psi^\prime(\third) \alpha C_A
- 30 \psi^\prime(\third) C_A
- 27 \alpha^2 C_A
- 8 \pi^2 \alpha C_A 
\right. \nonumber \\
&& \left. ~~~~~\,
- 108 \alpha C_A
+ 20 \pi^2 C_A 
- 669 C_A
+ 240 \Nf T_F \right] \frac{a}{216} ~+~ O(a^2) \nonumber \\
C_g^{\MOMqs}(a,\alpha) &=& 
1 + \left[ - \, 6 \psi^\prime(\third) \alpha^2 C_A
+ 24 \psi^\prime(\third) \alpha C_A
+ 96 \psi^\prime(\third) \alpha C_F
+ 78 \psi^\prime(\third) C_A
\right. \nonumber \\
&& \left. ~~~~~\,
- 48 \psi^\prime(\third) C_F
+ 4 \pi^2 \alpha^2 C_A 
+ 27 \alpha^2 C_A
- 16 \pi^2 \alpha C_A 
- 54 \alpha C_A
\right. \nonumber \\
&& \left. ~~~~~\,
- 64 \pi^2 \alpha C_F 
- 216 \alpha C_F
- 52 \pi^2 C_A 
- 993 C_A
+ 32 \pi^2 C_F 
+ 432 C_F
\right. \nonumber \\
&& \left. ~~~~~\,
+ 240 \Nf T_F \right] \frac{a}{216} ~+~ O(a^2) ~.
\end{eqnarray}
At one loop the gauge parameter mapping is the same in each scheme, similar to
the MAG, and thus we  have 
\begin{equation}
\alpha_{\MOMis} ~=~
\alpha + \left[ - \, 9 \alpha^2 C_A - 36 \alpha C_A - 97 C_A 
+ 80 \Nf T_F \right] \frac{\alpha a}{36} ~+~ O(a^2) ~.
\end{equation}
Equally the conversion functions for the wave function renormalization
constants are the same in each scheme and are
\begin{eqnarray}
C_A(a,\alpha) &=& 1 + \left[ 9 \alpha^2 C_A + 18 \alpha C_A + 97 C_A 
- 80 \Nf T_F \right] \frac{a}{36} ~+~ O(a^2) \nonumber \\
C_c(a,\alpha) &=& 1 + C_A a ~+~ O(a^2) \nonumber \\
C_\psi(a,\alpha) &=& 1 - \alpha C_F a ~+~ O(a^2) ~. 
\end{eqnarray} 

As before we have checked that the scheme independent one loop parts of the
renormalization group functions correctly emerge in our direct evaluation.
Equipped with these and the one loop conversion functions which derive from
the finite parts of the Green's functions we find the following results for
the renormalization group functions. First, for the MOMggg scheme we have  
\begin{eqnarray}
\beta^{\MOMgggs}(a,\alpha) &=&
\left[ - \, 11 C_A + 4 \Nf T_F \right] \frac{a^2}{3} \nonumber \\
&& + \left[ - \, 72 \psi^\prime(\third) \alpha^3 C_A^2
+ 786 \psi^\prime(\third) \alpha^2 C_A^2
- 384 \psi^\prime(\third) \alpha^2 C_A \Nf T_F
\right. \nonumber \\
&& \left. ~~~
- \, 1404 \psi^\prime(\third) \alpha C_A^2
+ 864 \psi^\prime(\third) \alpha C_A \Nf T_F
- 81 \alpha^4 C_A^2
+ 48 \pi^2 \alpha^3 C_A^2 
\right. \nonumber \\
&& \left. ~~~
+ 1026 \alpha^3 C_A^2
- 432 \alpha^3 C_A \Nf T_F
- 524 \pi^2 \alpha^2 C_A^2 
- 3051 \alpha^2 C_A^2
\right. \nonumber \\
&& \left. ~~~
+ 256 \pi^2 \alpha^2 C_A \Nf T_F
+ 1728 \alpha^2 C_A \Nf T_F
+ 936 \pi^2 \alpha C_A^2 
+ 2106 \alpha C_A^2
\right. \nonumber \\
&& \left. ~~~
- 576 \pi^2 \alpha C_A \Nf T_F
- 1296 \alpha C_A \Nf T_F
- 14688 C_A^2
+ 8640 C_A \Nf T_F
\right. \nonumber \\
&& \left. ~~~
+ 5184 C_F \Nf T_F \right] \frac{a^3}{1296} ~+~ O(a^4) \nonumber \\
\gamma_A^{\MOMgggs}(a,\alpha) &=&
\left[ 3 \alpha C_A - 13 C_A + 8 \Nf T_F \right] \frac{a}{6} \nonumber \\
&& + \left[ - \, 108 \psi^\prime(\third) \alpha^3 C_A^2
+ 954 \psi^\prime(\third) \alpha^2 C_A^2
- 288 \psi^\prime(\third) \alpha^2 C_A \Nf T_F
\right. \nonumber \\
&& \left. ~~~
- 2520 \psi^\prime(\third) \alpha C_A^2
+ 2448 \psi^\prime(\third) \alpha C_A \Nf T_F
+ 1794 \psi^\prime(\third) C_A^2
\right. \nonumber \\
&& \left. ~~~
- 6096 \psi^\prime(\third) C_A \Nf T_F
+ 3072 \psi^\prime(\third) \Nf^2 T_F^2
- 81 \alpha^4 C_A^2
+ 72 \pi^2 \alpha^3 C_A^2 
\right. \nonumber \\
&& \left. ~~~
+ 837 \alpha^3 C_A^2
- 216 \alpha^3 C_A \Nf T_F
- 636 \pi^2 \alpha^2 C_A^2 
- 1539 \alpha^2 C_A^2
\right. \nonumber \\
&& \left. ~~~
+ 192 \pi^2 \alpha^2 C_A \Nf T_F
+ 648 \alpha^2 C_A \Nf T_F
+ 1680 \pi^2 \alpha C_A^2 
- 135 \alpha C_A^2
\right. \nonumber \\
&& \left. ~~~
- 1632 \pi^2 \alpha C_A \Nf T_F
- 1512 \alpha C_A \Nf T_F
- 1196 \pi^2 C_A^2 
- 2655 C_A^2
\right. \nonumber \\
&& \left. ~~~
+ 4064 \pi^2 C_A \Nf T_F
+ 2304 C_A \Nf T_F
+ 7776 C_F \Nf T_F
- 2048 \pi^2 \Nf^2 T_F^2
\right. \nonumber \\
&& \left. ~~~
+ 1152 \Nf^2 T_F^2 \right] \frac{a^2}{1944} ~+~ O(a^3) \nonumber \\ 
\gamma_\alpha^{\MOMgggs}(a,\alpha) &=&
\left[ - \, 3 \alpha C_A + 26 C_A - 16 \Nf T_F \right] \frac{a}{12} \nonumber \\
&& + \left[ 108 \psi^\prime(\third) \alpha^3 C_A^2
- 1422 \psi^\prime(\third) \alpha^2 C_A^2
+ 576 \psi^\prime(\third) \alpha^2 C_A \Nf T_F
\right. \nonumber \\
&& \left. ~~~
+ 4626 \psi^\prime(\third) \alpha C_A^2
- 3744 \psi^\prime(\third) \alpha C_A \Nf T_F
- 3588 \psi^\prime(\third) C_A^2
\right. \nonumber \\
&& \left. ~~~
+ 12192 \psi^\prime(\third) C_A \Nf T_F
- 6144 \psi^\prime(\third) \Nf^2 T_F^2
+ 81 \alpha^4 C_A^2
- 72 \pi^2 \alpha^3 C_A^2 
\right. \nonumber \\
&& \left. ~~~
- 945 \alpha^3 C_A^2
+ 432 \alpha^3 C_A \Nf T_F
+ 948 \pi^2 \alpha^2 C_A^2 
+ 4050 \alpha^2 C_A^2
\right. \nonumber \\
&& \left. ~~~
- 384 \pi^2 \alpha^2 C_A \Nf T_F
- 1296 \alpha^2 C_A \Nf T_F
- 3084 \pi^2 \alpha C_A^2 
- 108 \alpha C_A^2
\right. \nonumber \\
&& \left. ~~~
+ 2496 \pi^2 \alpha C_A \Nf T_F
+ 3456 \alpha C_A \Nf T_F
+ 2392 \pi^2 C_A^2 
+ 5310 C_A^2
\right. \nonumber \\
&& \left. ~~~
- 8128 \pi^2 C_A \Nf T_F
- 4608 C_A \Nf T_F
- 15552 C_F \Nf T_F
+ 4096 \pi^2 \Nf^2 T_F^2
\right. \nonumber \\
&& \left. ~~~
- 2304 \Nf^2 T_F^2 \right] \frac{a^2}{3888} ~+~ O(a^3) \nonumber \\
\gamma_c^{\MOMgggs}(a,\alpha) &=&
[ \alpha - 3 ] \frac{C_A a}{4} \nonumber \\
&& + \left[ - \, 36 \psi^\prime(\third) \alpha^3 C_A
+ 270 \psi^\prime(\third) \alpha^2 C_A
- 624 \psi^\prime(\third) \alpha C_A
+ 384 \psi^\prime(\third) \alpha \Nf T_F
\right. \nonumber \\
&& \left. ~~~
+ 414 \psi^\prime(\third) C_A
- 1152 \psi^\prime(\third) \Nf T_F
- 27 \alpha^4 C_A
+ 24 \pi^2 \alpha^3 C_A 
+ 324 \alpha^3 C_A
\right. \nonumber \\
&& \left. ~~~
- 180 \pi^2 \alpha^2 C_A 
- 324 \alpha^2 C_A
+ 416 \pi^2 \alpha C_A 
- 855 \alpha C_A
- 256 \pi^2 \alpha \Nf T_F
\right. \nonumber \\
&& \left. ~~~
+ 144 \alpha \Nf T_F
- 276 \pi^2 C_A 
- 189 C_A
+ 768 \pi^2 \Nf T_F
+ 216 \Nf T_F \right] \frac{C_A a^2}{1296} \nonumber \\
&& +~ O(a^3) \nonumber \\
\gamma_\psi^{\MOMgggs}(a,\alpha) &=&
\alpha C_F a \nonumber \\
&& + \left[ - \, 36 \psi^\prime(\third) \alpha^3 C_A
+ 162 \psi^\prime(\third) \alpha^2 C_A
- 138 \psi^\prime(\third) \alpha C_A
+ 384 \psi^\prime(\third) \alpha \Nf T_F
\right. \nonumber \\
&& \left. ~~~
- 27 \alpha^4 C_A
+ 24 \pi^2 \alpha^3 C_A 
+ 243 \alpha^3 C_A
- 108 \pi^2 \alpha^2 C_A 
+ 162 \alpha^2 C_A
\right. \nonumber \\
&& \left. ~~~
+ 92 \pi^2 \alpha C_A 
- 369 \alpha C_A
- 256 \pi^2 \alpha \Nf T_F
+ 144 \alpha \Nf T_F
+ 2025 C_A
\right. \nonumber \\
&& \left. ~~~
- 486 C_F
- 648 \Nf T_F \right] \frac{C_F a^2}{324} ~+~ O(a^3) ~. 
\end{eqnarray} 
The results for the scheme based on the ghost vertex are similar since 
\begin{eqnarray}
\beta^{\MOMhs}(a,\alpha) &=&
\left[ - \, 11 C_A + 4 \Nf T_F \right] \frac{a^2}{3} \nonumber \\
&& + \left[ 18 \psi^\prime(\third) \alpha^2 C_A^2
- 156 \psi^\prime(\third) \alpha C_A^2
+ 96 \psi^\prime(\third) \alpha C_A \Nf T_F
- 81 \alpha^3 C_A^2
\right. \nonumber \\
&& \left. ~~~
- 12 \pi^2 \alpha^2 C_A^2
+ 540 \alpha^2 C_A^2
- 432 \alpha^2 C_A \Nf T_F
+ 104 \pi^2 \alpha C_A^2
+ 1404 \alpha C_A^2
\right. \nonumber \\
&& \left. ~~~
- 64 \pi^2 \alpha C_A \Nf T_F
- 864 \alpha C_A \Nf T_F
- 7344 C_A^2
+ 4320 C_A \Nf T_F
\right. \nonumber \\
&& \left. ~~~
+ 2592 C_F \Nf T_F \right] \frac{a^3}{648} ~+~ O(a^4) \nonumber \\
\gamma_A^{\MOMhs}(a,\alpha) &=&
\left[ 3 \alpha C_A - 13 C_A + 8 \Nf T_F \right] \frac{a}{6} \nonumber \\
&& + \left[ 36 \psi^\prime(\third) \alpha^2 C_A^2
- 246 \psi^\prime(\third) \alpha C_A^2
+ 96 \psi^\prime(\third) \alpha C_A \Nf T_F
+ 390 \psi^\prime(\third) C_A^2
\right. \nonumber \\
&& \left. ~~~
- 240 \psi^\prime(\third) C_A \Nf T_F
- 81 \alpha^3 C_A^2
- 24 \pi^2 \alpha^2 C_A^2
+ 459 \alpha^2 C_A^2
- 432 \alpha^2 C_A \Nf T_F
\right. \nonumber \\
&& \left. ~~~
+ 164 \pi^2 \alpha C_A^2
+ 675 \alpha C_A^2
- 64 \pi^2 \alpha C_A \Nf T_F
- 864 \alpha C_A \Nf T_F
- 260 \pi^2 C_A^2
\right. \nonumber \\
&& \left. ~~~
- 2484 C_A^2
+ 160 \pi^2 C_A \Nf T_F
+ 2376 C_A \Nf T_F
\right. \nonumber \\
&& \left. ~~~
+ 2592 C_F \Nf T_F \right] \frac{a^2}{648} ~+~ O(a^3) \nonumber \\
\gamma_\alpha^{\MOMhs}(a,\alpha) &=&
\left[ - \, 3 \alpha C_A + 26 C_A - 16 \Nf T_F \right] \frac{a}{12} 
\nonumber \\
&& + \left[ - \, 36 \psi^\prime(\third) \alpha^2 C_A^2
+ 402 \psi^\prime(\third) \alpha C_A^2
- 192 \psi^\prime(\third) \alpha C_A \Nf T_F
- 780 \psi^\prime(\third) C_A^2
\right. \nonumber \\
&& \left. ~~~
+ 480 \psi^\prime(\third) C_A \Nf T_F
+ 162 \alpha^3 C_A^2
+ 24 \pi^2 \alpha^2 C_A^2
- 675 \alpha^2 C_A^2
\right. \nonumber \\
&& \left. ~~~
+ 864 \alpha^2 C_A \Nf T_F
- 268 \pi^2 \alpha C_A^2
- 1107 \alpha C_A^2
+ 128 \pi^2 \alpha C_A \Nf T_F
\right. \nonumber \\
&& \left. ~~~
+ 1728 \alpha C_A \Nf T_F
+ 520 \pi^2 C_A^2
+ 4968 C_A^2
- 320 \pi^2 C_A \Nf T_F
\right. \nonumber \\
&& \left. ~~~
- 4752 C_A \Nf T_F
- 5184 C_F \Nf T_F \right] \frac{a^2}{1296} ~+~ O(a^3) \nonumber \\
\gamma_c^{\MOMhs}(a,\alpha) &=&
[ \alpha - 3 ] \frac{C_A a}{4} \nonumber \\
&& + \left[ 12 \psi^\prime(\third) \alpha^2 C_A
- 66 \psi^\prime(\third) \alpha C_A
+ 90 \psi^\prime(\third) C_A
- 8 \pi^2 \alpha^2 C_A
+ 108 \alpha^2 C_A
\right. \nonumber \\
&& \left. ~~~
+ 44 \pi^2 \alpha C_A
- 81 \alpha C_A
- 60 \pi^2 C_A
- 432 C_A
+ 216 \Nf T_F \right] \frac{C_A a^2}{432} ~+~ O(a^3) \nonumber \\
\gamma_\psi^{\MOMhs}(a,\alpha) &=&
\alpha C_F a \nonumber \\
&& + \left[ 12 \psi^\prime(\third) \alpha^2 C_A
- 30 \psi^\prime(\third) \alpha C_A
- 8 \pi^2 \alpha^2 C_A
+ 27 \alpha^2 C_A
+ 20 \pi^2 \alpha C_A
\right. \nonumber \\
&& \left. ~~~
+ 675 C_A
- 162 C_F
- 216 \Nf T_F \right] \frac{C_F a^2}{108} ~+~ O(a^3) ~.
\end{eqnarray}
Finally for the MOMq scheme the results are more involved since 
\begin{eqnarray}
\beta^{\MOMqs}(a,\alpha) &=&
\left[ - \, 11 C_A + 4 \Nf T_F \right] \frac{a^2}{3} \nonumber \\
&& + \left[ - \, 18 \psi^\prime(\third) \alpha^3 C_A^2
+ 192 \psi^\prime(\third) \alpha^2 C_A^2
+ 144 \psi^\prime(\third) \alpha^2 C_A C_F
\right. \nonumber \\
&& \left. ~~~
- 96 \psi^\prime(\third) \alpha^2 C_A \Nf T_F
- 312 \psi^\prime(\third) \alpha C_A^2
- 1248 \psi^\prime(\third) \alpha C_A C_F
\right. \nonumber \\
&& \left. ~~~
+ 192 \psi^\prime(\third) \alpha C_A \Nf T_F
+ 768 \psi^\prime(\third) \alpha C_F \Nf T_F
+ 12 \pi^2 \alpha^3 C_A^2
+ 81 \alpha^3 C_A^2
\right. \nonumber \\
&& \left. ~~~
- 128 \pi^2 \alpha^2 C_A^2
- 783 \alpha^2 C_A^2
- 96 \pi^2 \alpha^2 C_A C_F
- 324 \alpha^2 C_A C_F
\right. \nonumber \\
&& \left. ~~~
+ 64 \pi^2 \alpha^2 C_A \Nf T_F
+ 432 \alpha^2 C_A \Nf T_F
+ 208 \pi^2 \alpha C_A^2
+ 702 \alpha C_A^2
\right. \nonumber \\
&& \left. ~~~
+ 832 \pi^2 \alpha C_A C_F
+ 2808 \alpha C_A C_F
- 128 \pi^2 \alpha C_A \Nf T_F
- 432 \alpha C_A \Nf T_F
\right. \nonumber \\
&& \left. ~~~
- 512 \pi^2 \alpha C_F \Nf T_F
- 1728 \alpha C_F \Nf T_F
- 7344 C_A^2
+ 4320 C_A \Nf T_F
\right. \nonumber \\
&& \left. ~~~
+ 2592 C_F \Nf T_F \right] \frac{a^3}{648} ~+~ O(a^4) \nonumber \\
\gamma_A^{\MOMqs}(a,\alpha) &=&
\left[ 3 \alpha C_A - 13 C_A + 8 \Nf T_F \right] \frac{a}{6} \nonumber \\
&& + \left[ - \, 18 \psi^\prime(\third) \alpha^3 C_A^2
+ 150 \psi^\prime(\third) \alpha^2 C_A^2
+ 288 \psi^\prime(\third) \alpha^2 C_A C_F
\right. \nonumber \\
&& \left. ~~~
- 48 \psi^\prime(\third) \alpha^2 C_A \Nf T_F
- 78 \psi^\prime(\third) \alpha C_A^2
- 1392 \psi^\prime(\third) \alpha C_A C_F
\right. \nonumber \\
&& \left. ~~~
+ 192 \psi^\prime(\third) \alpha C_A \Nf T_F
+ 768 \psi^\prime(\third) \alpha C_F \Nf T_F
- 1014 \psi^\prime(\third) C_A^2
\right. \nonumber \\
&& \left. ~~~
+ 624 \psi^\prime(\third) C_A C_F
+ 624 \psi^\prime(\third) C_A \Nf T_F
- 384 \psi^\prime(\third) C_F \Nf T_F
\right. \nonumber \\
&& \left. ~~~
+ 12 \pi^2 \alpha^3 C_A^2
+ 81 \alpha^3 C_A^2
- 100 \pi^2 \alpha^2 C_A^2
- 81 \alpha^2 C_A^2
- 192 \pi^2 \alpha^2 C_A C_F
\right. \nonumber \\
&& \left. ~~~
- 648 \alpha^2 C_A C_F
+ 32 \pi^2 \alpha^2 C_A \Nf T_F
+ 52 \pi^2 \alpha C_A^2
- 999 \alpha C_A^2
+ 928 \pi^2 \alpha C_A C_F
\right. \nonumber \\
&& \left. ~~~
+ 4104 \alpha C_A C_F
- 128 \pi^2 \alpha C_A \Nf T_F
- 432 \alpha C_A \Nf T_F
- 512 \pi^2 \alpha C_F \Nf T_F
\right. \nonumber \\
&& \left. ~~~
- 1728 \alpha C_F \Nf T_F
+ 676 \pi^2 C_A^2
+ 1728 C_A^2
- 416 \pi^2 C_A C_F
- 5616 C_A C_F
\right. \nonumber \\
&& \left. ~~~
- 416 \pi^2 C_A \Nf T_F
- 216 C_A \Nf T_F
+ 256 \pi^2 C_F \Nf T_F
+ 6048 C_F \Nf T_F \right] \frac{a^2}{648} \nonumber \\
&& +~ O(a^3) \nonumber \\
\gamma_\alpha^{\MOMqs}(a,\alpha) &=&
\left[ - \, 3 \alpha C_A + 26 C_A - 16 \Nf T_F \right] \frac{a}{12} 
\nonumber \\
&& + \left[ 18 \psi^\prime(\third) \alpha^3 C_A^2
- 228 \psi^\prime(\third) \alpha^2 C_A^2
- 288 \psi^\prime(\third) \alpha^2 C_A C_F
+ 96 \psi^\prime(\third) \alpha^2 C_A \Nf T_F
\right. \nonumber \\
&& \left. ~~~
+ 390 \psi^\prime(\third) \alpha C_A^2
+ 2640 \psi^\prime(\third) \alpha C_A C_F
- 384 \psi^\prime(\third) \alpha C_A \Nf T_F
\right. \nonumber \\
&& \left. ~~~
- 1536 \psi^\prime(\third) \alpha C_F \Nf T_F
+ 2028 \psi^\prime(\third) C_A^2
- 1248 \psi^\prime(\third) C_A C_F
\right. \nonumber \\
&& \left. ~~~
- 1248 \psi^\prime(\third) C_A \Nf T_F
+ 768 \psi^\prime(\third) C_F \Nf T_F
- 12 \pi^2 \alpha^3 C_A^2
+ 152 \pi^2 \alpha^2 C_A^2
\right. \nonumber \\
&& \left. ~~~
+ 567 \alpha^2 C_A^2
+ 192 \pi^2 \alpha^2 C_A C_F
+ 648 \alpha^2 C_A C_F
- 64 \pi^2 \alpha^2 C_A \Nf T_F
\right. \nonumber \\
&& \left. ~~~
- 260 \pi^2 \alpha C_A^2
+ 1269 \alpha C_A^2
- 1760 \pi^2 \alpha C_A C_F
- 6912 \alpha C_A C_F
\right. \nonumber \\
&& \left. ~~~
+ 256 \pi^2 \alpha C_A \Nf T_F
+ 864 \alpha C_A \Nf T_F
+ 1024 \pi^2 \alpha C_F \Nf T_F
+ 3456 \alpha C_F \Nf T_F
\right. \nonumber \\
&& \left. ~~~
- 1352 \pi^2 C_A^2 
- 3456 C_A^2
+ 832 \pi^2 C_A C_F 
+ 11232 C_A C_F
\right. \nonumber \\
&& \left. ~~~
+ 832 \pi^2 C_A \Nf T_F
+ 432 C_A \Nf T_F
- 512 \pi^2 C_F \Nf T_F
- 12096 C_F \Nf T_F \right] \frac{a^2}{1296} \nonumber \\
&& +~ O(a^3) \nonumber \\ 
\gamma_c^{\MOMqs}(a,\alpha) &=&
[ \alpha - 3 ] \frac{C_A a}{4} \nonumber \\
&& + \left[ - \, 6 \psi^\prime(\third) \alpha^3 C_A
+ 42 \psi^\prime(\third) \alpha^2 C_A
+ 96 \psi^\prime(\third) \alpha^2 C_F
+ 6 \psi^\prime(\third) \alpha C_A
\right. \nonumber \\
&& \left. ~~~
- 336 \psi^\prime(\third) \alpha C_F
- 234 \psi^\prime(\third) C_A
+ 144 \psi^\prime(\third) C_F
+ 4 \pi^2 \alpha^3 C_A
+ 54 \alpha^3 C_A
\right. \nonumber \\
&& \left. ~~~
- 28 \pi^2 \alpha^2 C_A
- 64 \pi^2 \alpha^2 C_F
- 216 \alpha^2 C_F
- 4 \pi^2 \alpha C_A
- 567 \alpha C_A
+ 224 \pi^2 \alpha C_F
\right. \nonumber \\
&& \left. ~~~
+ 1080 \alpha C_F
+ 156 \pi^2 C_A
+ 540 C_A
- 96 \pi^2 C_F
- 1296 C_F
+ 216 \Nf T_F \right] \frac{a^2}{432} \nonumber \\
&& +~ O(a^3) \nonumber \\
\gamma_\psi^{\MOMqs}(a,\alpha) &=&
\alpha C_F a \nonumber \\
&& + \left[ - \, 6 \psi^\prime(\third) \alpha^3 C_A
+ 24 \psi^\prime(\third) \alpha^2 C_A
+ 96 \psi^\prime(\third) \alpha^2 C_F
+ 78 \psi^\prime(\third) \alpha C_A
\right. \nonumber \\
&& \left. ~~~
- 48 \psi^\prime(\third) \alpha C_F
+ 4 \pi^2 \alpha^3 C_A
+ 54 \alpha^3 C_A
- 16 \pi^2 \alpha^2 C_A
+ 81 \alpha^2 C_A
\right. \nonumber \\
&& \left. ~~~
- 64 \pi^2 \alpha^2 C_F
- 216 \alpha^2 C_F
- 52 \pi^2 \alpha C_A
- 324 \alpha C_A
+ 32 \pi^2 \alpha C_F
\right. \nonumber \\
&& \left. ~~~
+ 432 \alpha C_F
+ 675 C_A
- 162 C_F
- 216 \Nf T_F \right] \frac{C_F a^2}{108} ~+~ O(a^3) ~.
\end{eqnarray} 
We have concentrated on the renormalization group functions for the 
Curci-Ferrari gauge. The explicit form of the various amplitudes can be deduced
from the MAG expressions given in the data file in the 
$\Nda/\Noda$~$\rightarrow$~$0$ limit. We have checked that these agree with the
direct evaluation performed in the Curci-Ferrari gauge itself.

\sect{$\Lambda$ parameters.}

Having provided all the one loop structure for the MAG and the Curci-Ferrari
gauge for the MOM schemes we now briefly discuss the relation between the
$\Lambda$ parameters in the MOM schemes to those in the $\MSbar$ scheme. This
parameter sets the fundamental scale in QCD and corresponds to the boundary
between infrared and ultraviolet physics. However, its actual value depends on
the renormalization scheme one is considering. Though one remarkable feature of
this non-perturbative quantity is that the ratio between $\Lambda$ parameters in
different schemes can be determined exactly from a {\em one} loop computation. 
In \cite{18} those relations for the various MOM schemes were determined and we
repeat that analysis here for the MAG and Curci-Ferrari gauges. In essence the 
ratio of parameters reflects the first term of the coupling constant conversion
function. First, we define
\begin{equation}
\frac{\Lambda^{\MOMis}}{\Lambda^{\MSbars}} ~=~ \exp \left[ 
\frac{\lambda^{\MOMis}(\alpha,\Nf)}{b_0} \right]
\end{equation}
where
\begin{equation}
b_0 ~=~ \frac{22}{3} C_A ~-~ \frac{8}{3} T_F \Nf
\end{equation}
originates from the one loop $\beta$-function. Then for each of the three MOM 
schemes in the MAG we have
\begin{eqnarray}
\lambda^{\MOMgggs}(\alpha,\Nf) &=& 
\frac{1}{324 {\Noda}} 
\left[ -~ 72 \psi^\prime(\third) \alpha^2 C_A {\Nda}
+ 36 \psi^\prime(\third) \alpha^2 C_A {\Noda}
+ 90 \psi^\prime(\third) \alpha C_A {\Nda}
\right. \nonumber \\
&& \left. ~~~~~~~~~~
- 162 \psi^\prime(\third) \alpha C_A {\Noda}
- 702 \psi^\prime(\third) C_A {\Nda}
+ 138 \psi^\prime(\third) C_A {\Noda}
\right. \nonumber \\
&& \left. ~~~~~~~~~~
- 384 \psi^\prime(\third) \Nf {\Noda} T_F
- 81 \alpha^3 C_A {\Nda}
+ 27 \alpha^3 C_A {\Noda}
+ 48 \pi^2 \alpha^2 C_A {\Nda}
\right. \nonumber \\
&& \left. ~~~~~~~~~~
+ 324 \alpha^2 C_A {\Nda}
- 24 \pi^2 \alpha^2 C_A {\Noda}
- 162 \alpha^2 C_A {\Noda}
- 60 \pi^2 \alpha C_A {\Nda}
\right. \nonumber \\
&& \left. ~~~~~~~~~~
- 243 \alpha C_A {\Nda}
+ 108 \pi^2 \alpha C_A {\Noda}
+ 243 \alpha C_A {\Noda}
+ 468 \pi^2 C_A {\Nda}
\right. \nonumber \\
&& \left. ~~~~~~~~~~
- 92 \pi^2 C_A {\Noda}
+ 2376 C_A {\Noda}
+ 256 \pi^2 \Nf {\Noda} T_F
- 864 \Nf {\Noda} T_F \right] \nonumber \\ 
\lambda^{\MOMhs}(\alpha,\Nf) &=& 
\frac{1}{108 {\Noda}} 
\left[ 
36 \psi^\prime(\third) \alpha C_A {\Nda}
- 12 \psi^\prime(\third) \alpha C_A {\Noda}
- 66 \psi^\prime(\third) C_A {\Nda}
\right. \nonumber \\
&& \left. ~~~~~~~~~~
+ 30 \psi^\prime(\third) C_A {\Noda}
- 54 \alpha^2 C_A {\Nda}
+ 27 \alpha^2 C_A {\Noda}
- 24 \pi^2 \alpha C_A {\Nda} 
\right. \nonumber \\
&& \left. ~~~~~~~~~~
- 108 \alpha C_A {\Nda}
+ 8 \pi^2 \alpha C_A {\Noda} 
+ 108 \alpha C_A {\Noda}
+ 44 \pi^2 C_A {\Nda} 
\right. \nonumber \\
&& \left. ~~~~~~~~~~
+ 162 C_A {\Nda}
- 20 \pi^2 C_A {\Noda} 
+ 669 C_A {\Noda}
- 240 \Nf {\Noda} T_F \right] \nonumber \\
\lambda^{\MOMqs}(\alpha,\Nf) &=& 
\frac{1}{108 \NF {\Noda}} 
\left[
-~ 12 \psi^\prime(\third) \alpha^2 C_A \NF {\Nda}
+ 6 \psi^\prime(\third) \alpha^2 C_A \NF {\Noda}
\right. \nonumber \\
&& \left. ~~~~~~~~~~~~~~
+ 24 \psi^\prime(\third) \alpha C_A \NF {\Nda}
- 24 \psi^\prime(\third) \alpha C_A \NF {\Noda}
- 96 \psi^\prime(\third) \alpha {\Noda}^2 T_F
\right. \nonumber \\
&& \left. ~~~~~~~~~~~~~~
- 60 \psi^\prime(\third) C_A \NF {\Nda}
- 78 \psi^\prime(\third) C_A \NF {\Noda}
+ 48 \psi^\prime(\third) C_F \NF {\Noda}
\right. \nonumber \\
&& \left. ~~~~~~~~~~~~~~
+ 8 \pi^2 \alpha^2 C_A \NF {\Nda}
+ 54 \alpha^2 C_A \NF {\Nda}
- 4 \pi^2\alpha^2 C_A \NF {\Noda}
\right. \nonumber \\
&& \left. ~~~~~~~~~~~~~~
- 27 \alpha^2 C_A \NF {\Noda}
- 16 \pi^2 \alpha C_A \NF {\Nda}
- 54 \alpha C_A \NF {\Nda}
\right. \nonumber \\
&& \left. ~~~~~~~~~~~~~~
+ 16 \pi^2 \alpha C_A \NF {\Noda}
+ 54 \alpha C_A \NF {\Noda}
+ 64 \pi^2 \alpha {\Noda}^2 T_F
\right. \nonumber \\
&& \left. ~~~~~~~~~~~~~~
+ 216 \alpha {\Noda}^2 T_F
+ 40 \pi^2 C_A \NF {\Nda}
+ 52 \pi^2 C_A \NF {\Noda}
\right. \nonumber \\
&& \left. ~~~~~~~~~~~~~~
+ 993 C_A \NF {\Noda}
- 32 \pi^2 C_F \NF {\Noda}
- 432 C_F \NF {\Noda}
\right. \nonumber \\
&& \left. ~~~~~~~~~~~~~~
- 240 \NF \Nf {\Noda} T_F
\right] ~. 
\end{eqnarray} 
While these are the explicit results it is perhaps more instructive to compare
with the Landau gauge results of \cite{18}. Therefore we have provided the
values for the same choice of $\Nf$ and $\alpha$ given in \cite{18} for each 
scheme for $SU(3)$ in Table $1$. Though it is important to note that our 
$\alpha$ is not the same parameter as in \cite{18} and also the distinction 
between the MS and $\MSbar$ results of \cite{18}. Interestingly for certain 
choices of $\alpha$ and $\Nf$ the ratio is less than unity.

{\begin{table}[ht]
\begin{center}
\begin{tabular}{|c||c||c|c|c|}
\hline
$\alpha$ & $\Nf$ & $\MOMggg$ & $\MOMh$ & $\MOMq$ \\
\hline
0 & 0 & 2.3583 & 2.5816 & 1.9562 \\
0 & 1 & 2.1127 & 2.6008 & 1.9359 \\
0 & 2 & 1.8642 & 2.6228 & 1.9129 \\
0 & 3 & 1.6167 & 2.6484 & 1.8869 \\
0 & 4 & 1.3668 & 2.6784 & 1.8572 \\
0 & 5 & 1.1239 & 2.7140 & 1.8229 \\
1 & 0 & 2.0664 & 2.8596 & 1.8073 \\
1 & 3 & 1.3739 & 3.0010 & 1.7128 \\
1 & 4 & 1.1480 & 3.0655 & 1.6729 \\
1 & 5 & 0.9298 & 3.1429 & 1.6271 \\
3 & 3 & 0.9591 & 4.1883 & 1.3858 \\
3 & 4 & 0.7787 & 4.3939 & 1.3308 \\
-2 & 4 & 1.8624 & 2.2372 & 2.2445 \\
\hline
\end{tabular}
\end{center}
\begin{center}
{Table $1$. Values of $\lambda^{\MOMis}(\alpha,\Nf)$ for the MAG in $SU(3)$.}
\end{center}
\end{table}}
We have repeated the analysis for the Curci-Ferrari gauge and the parallel
results are presented in Table $2$. Those for the $\MOMggg$ and $\MOMq$ schemes
are equivalent to those of the linear covariant gauge fixing of \cite{18}. This
is because the coupling constant mapping is the same for both cases despite the
fact that the ghost-gluon vertex is different. This does not affect the one
loop vertices since the differences cancel out. However, this is not the case 
for the $\MOMh$ scheme since the quartic ghost vertex contributes to the 
mapping for all $\alpha$ and in the Landau gauge case the differences in the 
ghost-gluon vertex are significant. However, the same increase and decrease of 
the ratio with $\alpha$ and $\Nf$ is parallel to that for the standard linear 
covariant gauge fixing results of \cite{18}.

{\begin{table}[ht]
\begin{center}
\begin{tabular}{|c||c||c|c|c|}
\hline
$\alpha$ & $\Nf$ & $\MOMggg$ & $\MOMh$ & $\MOMq$ \\
\hline
0 & 0 & 3.3341 & 2.6588 & 2.1379 \\
0 & 1 & 3.0543 & 2.6837 & 2.1277 \\
0 & 2 & 2.7644 & 2.7123 & 2.1163 \\
0 & 3 & 2.4654 & 2.7456 & 2.1032 \\
0 & 4 & 2.1587 & 2.7846 & 2.0881 \\
0 & 5 & 1.8471 & 2.8312 & 2.0706 \\
1 & 0 & 2.8957 & 2.9893 & 1.9075 \\
1 & 3 & 2.0751 & 3.1684 & 1.8296 \\
1 & 4 & 1.7921 & 3.2505 & 1.7964 \\
1 & 5 & 1.5088 & 3.3496 & 1.7581 \\
3 & 3 & 1.8392 & 5.4177 & 1.3110 \\
3 & 4 & 1.5732 & 5.8018 & 1.2533 \\
-2 & 4 & 2.5437 & 2.6772 & 2.6597 \\
\hline
\end{tabular}
\end{center}
\begin{center}
{Table $2$. Values of $\lambda^{\MOMis}(\alpha,\Nf)$ for the Curci-Ferrari gauge
in $SU(3)$.}
\end{center}
\end{table}}

\sect{Discussion.}

We make some comments on our analysis. First, we have provided all the 
information on the $3$-point vertex functions relevant for the definition of
the MOM schemes for the maximal abelian gauge. This is an analysis parallel to
that of \cite{18} for QCD fixed in the canonical linear covariant gauge. One
motivation was to provide this data in relation to future lattice analyses of
the vertex functions in the infrared in order to have precision matching at
high energy. Moreover, the explicit values of the amplitudes will be useful for
assisting overlap with Schwinger-Dyson studies. Several features which were 
observed in \cite{30} are again present. One is the relation to the 
Curci-Ferrari gauge in that results from the latter can be derived from the MAG
in the replica-like limit where the centre of the group is formally excluded. 
However, we have verified that the results in this limit are consistent with 
the direct calculation in the Curci-Ferrari gauge itself. Given properties of 
the renormalization group equation the one loop conversion functions for 
relating parameters in the MOM schemes to those of the $\MSbar$ scheme have 
allowed us to compute the {\em two} loop renormalization group functions in 
each of the three MOM schemes. These have direct parallels with those of 
\cite{30} since they are based on the triple gluon, ghost-gluon and quark-gluon
vertices. Though an essential difference here is that with the split nature of 
the colour group in the MAG, it is the vertices with the off-diagonal gluons 
which are relevant. This is due in part to the fact that there are 
Slavnov-Taylor identities which ensure that the structure with vertices with 
centre gluons are predetermined. Indeed this is not unrelated to the fact these
gluons are similar to the background fields of the background field gauge of 
\cite{56,57,58,59,60,61} with the off-diagonal gluons corresponding to the 
quantum fluctuations. Whether this scenario is significant in the picture of 
abelian monopoles underlying a picture of colour confinement would be 
interesting to investigate.

\vspace{1cm}
\noindent
{\bf Acknowledgements.} This work was carried out with the support of a John
Lennon Memorial Scholarship and an STFC studentship (JMB).

\appendix

\sect{Tensor basis.}

In this appendix we record for completeness the tensor basis for each of the 
three $3$-point vertices, using the same notation as \cite{19}. For the triple
gluon vertex we use the original tensor basis with the basis tensors
\begin{eqnarray}
{\cal P}^{\mbox{\footnotesize{ggg}}}_{(1) \mu \nu \sigma }(p,q) &=&
\eta_{\mu \nu} p_\sigma ~~,~~
{\cal P}^{\mbox{\footnotesize{ggg}}}_{(2) \mu \nu \sigma }(p,q) ~=~
\eta_{\nu \sigma} p_\mu ~~,~~
{\cal P}^{\mbox{\footnotesize{ggg}}}_{(3) \mu \nu \sigma }(p,q) ~=~
\eta_{\sigma \mu} p_\nu \nonumber \\
{\cal P}^{\mbox{\footnotesize{ggg}}}_{(4) \mu \nu \sigma }(p,q) &=&
\eta_{\mu \nu} q_\sigma ~~,~~
{\cal P}^{\mbox{\footnotesize{ggg}}}_{(5) \mu \nu \sigma }(p,q) ~=~
\eta_{\nu \sigma} q_\mu ~~,~~
{\cal P}^{\mbox{\footnotesize{ggg}}}_{(6) \mu \nu \sigma }(p,q) ~=~
\eta_{\sigma \mu} q_\nu \nonumber \\
{\cal P}^{\mbox{\footnotesize{ggg}}}_{(7) \mu \nu \sigma }(p,q) &=&
\frac{1}{\mu^2} p_\mu p_\nu p_\sigma ~~,~~
{\cal P}^{\mbox{\footnotesize{ggg}}}_{(8) \mu \nu \sigma }(p,q) ~=~
\frac{1}{\mu^2} p_\mu p_\nu q_\sigma ~~,~~
{\cal P}^{\mbox{\footnotesize{ggg}}}_{(9) \mu \nu \sigma }(p,q) ~=~
\frac{1}{\mu^2} p_\mu q_\nu p_\sigma \nonumber \\
{\cal P}^{\mbox{\footnotesize{ggg}}}_{(10) \mu \nu \sigma }(p,q) &=&
\frac{1}{\mu^2} q_\mu p_\nu p_\sigma ~~,~~
{\cal P}^{\mbox{\footnotesize{ggg}}}_{(11) \mu \nu \sigma }(p,q) ~=~
\frac{1}{\mu^2} p_\mu q_\nu q_\sigma ~~,~~
{\cal P}^{\mbox{\footnotesize{ggg}}}_{(12) \mu \nu \sigma }(p,q) ~=~
\frac{1}{\mu^2} q_\mu p_\nu q_\sigma \nonumber \\
{\cal P}^{\mbox{\footnotesize{ggg}}}_{(13) \mu \nu \sigma }(p,q) &=&
\frac{1}{\mu^2} q_\mu q_\nu p_\sigma ~~,~~
{\cal P}^{\mbox{\footnotesize{ggg}}}_{(14) \mu \nu \sigma }(p,q) ~=~
\frac{1}{\mu^2} q_\mu q_\nu q_\sigma ~.
\end{eqnarray}
For the associated projection matrix we partition it into submatrices for ease
of presentation. With the general form
\begin{eqnarray}
{\cal M}^{\mbox{\footnotesize{ggg}}} &=& -~ \frac{1}{27(d-2)} \left(
\begin{array}{ccc}
{\cal M}^{\mbox{\footnotesize{ggg}}}_{11} &
{\cal M}^{\mbox{\footnotesize{ggg}}}_{12} &
{\cal M}^{\mbox{\footnotesize{ggg}}}_{13} \\
{\cal M}^{\mbox{\footnotesize{ggg}}}_{21} &
{\cal M}^{\mbox{\footnotesize{ggg}}}_{22} &
{\cal M}^{\mbox{\footnotesize{ggg}}}_{23} \\
{\cal M}^{\mbox{\footnotesize{ggg}}}_{31} &
{\cal M}^{\mbox{\footnotesize{ggg}}}_{32} &
{\cal M}^{\mbox{\footnotesize{ggg}}}_{33} \\
\end{array}
\right) \nonumber
\end{eqnarray}
then each of the submatrices are
\begin{eqnarray}
{\cal M}^{\mbox{\footnotesize{ggg}}}_{11} &=& \left(
\begin{array}{cccccc}
36 & 0 & 0 & 18 & 0 & 0 \\
0 & 36 & 0 & 0 & 18 & 0 \\
0 & 0 & 36 & 0 & 0 & 18 \\
18 & 0 & 0 & 36 & 0 & 0 \\
0 & 18 & 0 & 0 & 36 & 0 \\
0 & 0 & 18 & 0 & 0 & 36 \\
\end{array}
\right) ~,~
{\cal M}^{\mbox{\footnotesize{ggg}}}_{12} ~=~ \left(
\begin{array}{cccc}
48 & 24 & 24 & 24 \\
48 & 24 & 24 & 24 \\
48 & 24 & 24 & 24 \\
24 & 48 & 12 & 12 \\
24 & 12 & 12 & 48 \\
24 & 12 & 48 & 12 \\
\end{array}
\right) \nonumber \\
{\cal M}^{\mbox{\footnotesize{ggg}}}_{13} &=& \left(
\begin{array}{cccc}
12 & 12 & 48 & 24 \\
48 & 12 & 12 & 24 \\
12 & 48 & 12 & 24 \\
24 & 24 & 24 & 48 \\
24 & 24 & 24 & 48 \\
24 & 24 & 24 & 48 \\
\end{array}
\right) ~,~
{\cal M}^{\mbox{\footnotesize{ggg}}}_{21} ~=~ \left(
\begin{array}{cccccc}
48 & 48 & 48 & 24 & 24 & 24 \\
24 & 24 & 24 & 48 & 12 & 12 \\
24 & 24 & 24 & 12 & 12 & 48 \\
24 & 24 & 24 & 12 & 48 & 12 \\
\end{array}
\right) \nonumber \\
{\cal M}^{\mbox{\footnotesize{ggg}}}_{22} &=& \left(
\begin{array}{cccc}
64 (d+1) & 32 (d+1) & 32 (d+1) & 32 (d+1) \\
32 (d+1) & 32 (2d-1) & 16 (d+1) & 16 (d+1) \\
32 (d+1) & 16 (d+1) & 32 (2d-1) & 16 (d+1) \\
32 (d+1) & 16 (d+1) & 16 (d+1) & 32 (2d-1) \\
\end{array}
\right) \nonumber \\
{\cal M}^{\mbox{\footnotesize{ggg}}}_{23} &=& \left(
\begin{array}{cccc}
16 (d+4) & 16 (d+4) & 16 (d+4) & 8 (d+10) \\
8 (4d+1) & 8 (4d+1) & 8 (d+4) & 16 (d+4) \\
8 (4d+1) & 8 (d+4) & 8 (4d+1) & 16 (d+4) \\
8 (d+4) & 8 (4d+1) & 8 (4d+1) & 16 (d+4) \\
\end{array}
\right) \nonumber \\
{\cal M}^{\mbox{\footnotesize{ggg}}}_{31} &=& \left(
\begin{array}{cccccc}
12 & 48 & 12 & 24 & 24 & 24 \\
12 & 12 & 48 & 24 & 24 & 24 \\
48 & 12 & 12 & 24 & 24 & 24 \\
24 & 24 & 24 & 48 & 48 & 48 \\
\end{array}
\right) \nonumber \\
{\cal M}^{\mbox{\footnotesize{ggg}}}_{32} &=& \left(
\begin{array}{cccc}
16 (d+4) & 8 (4d+1) & 8 (4d+1) & 8 (d+4) \\
16 (d+4) & 8 (4d+1) & 8 (d+4) & 8 (4d+1) \\
16 (d+4) & 8 (d+4) & 8 (4d+1) & 8 (4d+1) \\
8 (d+10) & 16 (d+4) & 16 (d+4) & 16 (d+4) \\
\end{array}
\right) \nonumber \\
{\cal M}^{\mbox{\footnotesize{ggg}}}_{33} &=& \left(
\begin{array}{cccc}
32 (2d-1) & 16 (d+1) & 16 (d+1) & 32 (d+1) \\
16 (d+1) & 32 (2d-1) & 16 (d+1) & 32 (d+1) \\
16 (d+1) & 16 (d+1) & 32 (2d-1) & 32 (d+1) \\
32 (d+1) & 32 (d+1) & 32 (d+1) & 64 (d+1) \\
\end{array}
\right) ~.
\end{eqnarray}
The situation for the remaining vertices is simple as the basis of each involve
fewer tensors. For the ghost-gluon vertex we have
\begin{equation}
{\cal P}^{\mbox{\footnotesize{ccg}}}_{(1) \sigma }(p,q) ~=~ p_\sigma ~~~,~~~
{\cal P}^{\mbox{\footnotesize{ccg}}}_{(2) \sigma }(p,q) ~=~ q_\sigma
\end{equation}
where
\begin{equation}
{\cal M}^{\mbox{\footnotesize{ccg}}} ~=~ -~ \frac{1}{3} \left(
\begin{array}{cc}
4 & 2 \\
2 & 4 \\
\end{array}
\right)
\end{equation}
is the projection matrix. Finally, the quark-gluon vertex basis is
\begin{eqnarray}
{\cal P}^{\mbox{\footnotesize{qqg}}}_{(1) \sigma }(p,q) &=&
\gamma_\sigma ~~~,~~~
{\cal P}^{\mbox{\footnotesize{qqg}}}_{(2) \sigma }(p,q) ~=~
\frac{{p}_\sigma \pslash}{\mu^2} ~~~,~~~
{\cal P}^{\mbox{\footnotesize{qqg}}}_{(3) \sigma }(p,q) ~=~
\frac{{p}_\sigma \qslash}{\mu^2} ~, \nonumber \\
{\cal P}^{\mbox{\footnotesize{qqg}}}_{(4) \sigma }(p,q) &=&
\frac{{q}_\sigma \pslash}{\mu^2} ~~~,~~~
{\cal P}^{\mbox{\footnotesize{qqg}}}_{(5) \sigma }(p,q) ~=~
\frac{{q}_\sigma \qslash}{\mu^2} ~~~,~~~
{\cal P}^{\mbox{\footnotesize{qqg}}}_{(6) \sigma }(p,q) ~=~
\frac{1}{\mu^2} \Gamma_{(3) \, \sigma p q} ~.
\end{eqnarray}
which leads to the projection matrix
\begin{equation}
{\cal M}^{\mbox{\footnotesize{qqg}}} ~=~ \frac{1}{36(d-2)} \left(
\begin{array}{cccccc}
9 & 12 & 6 & 6 & 12 & 0 \\
12 & 16 (d - 1) &  8 (d - 1) &  8 (d - 1) & 4 (d + 2) & 0 \\
6 & 8 (d - 1) & 4 (4 d - 7) &  4 (d - 1) & 8 (d - 1) & 0 \\
6 & 8 (d - 1) &  4 (d - 1) & 4 (4 d - 7) & 8 (d - 1) & 0 \\
12 & 4 (d + 2) &  8 (d - 1) &  8 (d - 1) & 16 (d - 1) & 0 \\
0 & 0 & 0 & 0 & 0 & - 12 \\
\end{array}
\right) ~.
\label{qqgm}
\end{equation}
We have used the convention that when a momentum is contracted with a Lorentz
index then that momentum appears instead of the index in the tensor.


\begin{thebibliography}{99}
\bibitem{1} S. Mandelstam, Phys. Rept. {\bf 23} (1976), 245.
\bibitem{2} Y. Nambu, Phys. Rev. {\bf D10} (1974), 4262. 
\bibitem{3} G. 't Hooft, High Energy Physics EPS Int. Conference,
Palermo 1975, ed. A. Zichichi.
\bibitem{4} G. 't Hooft, Nucl. Phys. {\bf B190} (1981), 455.
\bibitem{5} Z.F. Ezawa \& A. Iwazaki, Phys. Rev. {\bf D25} (1982), 2681.
\bibitem{6} A.S. Kronfeld, G. Schierholz \& U.J. Wiese, Nucl. Phys. {\bf B293}
(1987), 461.
\bibitem{7} A.S. Kronfeld, M.L. Laursen, G. Schierholz \& U.J. Wiese, Phys.
Lett. {\bf B198} (1987), 516.
\bibitem{8} M.Q. Huber, K. Schwenzer \& R. Alkofer, Eur. Phys. J. {\bf C68}
(2010), 581.
\bibitem{9} R. Alkofer, M.Q. Huber, V. Mader \& A. Windisch, PoS QCD-TNT-II
(2011), 003.
\bibitem{10} V. Mader \& R. Alkofer, PoS ConfinementX (2012), 063.
\bibitem{11} T. Suzuki \& I. Yotsuyanagi, Phys. Rev. {\bf D42} (1990), 4257.
\bibitem{12} S. Hioki, S. Kitahara, S. Kiura, Y. Matsubara, O. Miyamura, S. 
Ohno \& T. Suzuki, Phys. Lett. {\bf B272} (1991), 326; Phys. Lett. {\bf B281} 
(1992), 416.
\bibitem{13} A. Mihara, A. Cucchieri \& T. Mendes, PoS LATTICE2008 (2008), 243.
\bibitem{14} T. Mendes, A. Cucchieri, A. Maas \& A. Mihara, arXiv:0809.3741.
\bibitem{15} S. Gongyo, T. Iritani \& H. Suganuma, Phys. Rev. {\bf D86} (2012),
094018.
\bibitem{16} S. Gongyo \& H. Suganuma, Phys. Rev. {\bf D87} (2013), 074506.
\bibitem{17} A. K\i z\i lers\"{u}, D.B. Leinweber, J.-I. Skullerud \& A.G.
Williams, Eur. Phys. J. {\bf C50} (2007), 87.
\bibitem{18} W. Celmaster \& R.J. Gonsalves, Phys. Rev. {\bf D20} (1979), 1420.
\bibitem{19} J.A. Gracey, Phys. Rev. {\bf D84} (2011), 085011.
\bibitem{20} H. Min, T. Lee \& P.Y. Pac, Phys. Rev. {\bf D32} (1985), 440.
\bibitem{21} K.-I. Kondo \& T. Shinohara, Phys. Lett. {\bf B491} (2000), 263.  
\bibitem{22} A.R. Fazio, V.E.R. Lemes, M.S. Sarandy \& S.P. Sorella, Phys. Rev.
{\bf D64} (2001), 085003.
\bibitem{23} K.-I. Kondo \& T. Shinohara, Prog. Theor. Phys. {\bf 105} (2001),
649.
\bibitem{24} T. Shinohara, Mod. Phys. Lett. {\bf A18} (2003), 1398.
\bibitem{25} K.-I. Kondo, Phys. Lett. {\bf B514} (2001), 335.  
\bibitem{26} T. Shinohara, T. Imai \& K.-I. Kondo, Int. J. Mod. Phys. {\bf A18} 
(2003), 5733.
\bibitem{27} H. Sawayanagi, Phys. Rev. {\bf D67} (2003), 045002.
\bibitem{28} D. Dudal \& H. Verschelde, J. Phys. {\bf A36} (2003), 8507.
\bibitem{29} D. Dudal, J.A. Gracey, V.E.R. Lemes, M.S. Sarandy, R.F. Sobreiro, 
S.P. Sorella \& H.Verschelde, Phys. Rev. {\bf D70} (2004), 114038. 
\bibitem{30} J.A. Gracey, JHEP {\bf 0504} (2005), 012.
\bibitem{31} G. Curci \& R. Ferrari, Nuovo Cim. {\bf A32} (1976), 151. 
\bibitem{32} A. Cucchieri \& T. Mendes, PoS LAT2007 (2007), 297.
\bibitem{33} I.L. Bogolubsky, E.M. Ilgenfritz, M. M\"{u}ller-Preussker \& A.
Sternbeck, PoS LAT2007 (2007), 290.
\bibitem{34} A. Maas, Phys. Rev. {\bf D75} (2007), 116004.
\bibitem{35} A. Sternbeck, L. von Smekal, D.B. Leinweber \& A.G. Williams,
PoS LAT2007 (2007), 304.
\bibitem{36} I.L. Bogolubsky, E.M. Ilgenfritz, M. M\"{u}ller-Preussker \& A.
Sternbeck, Phys. Lett. {\bf B676} (2009), 69.
\bibitem{37} A. Cucchieri \& T. Mendes, Phys. Rev. Lett. {\bf 100} (2008),
241601.
\bibitem{38} A. Cucchieri \& T. Mendes, Phys. Rev. {\bf D78} (2008), 094503.
\bibitem{39} O. Oliveira \& P.J. Silva, Phys. Rev. {\bf D79} (2009), 031501.
\bibitem{40} Ph. Boucaud, J.P. Leroy, A.L. Yaounac, J. Micheli, O. P\`{e}ne \&
J. Rodr\'{\i}guez-Quintero, JHEP {\bf 0806} (2008), 099.
\bibitem{41} G. 't Hooft \& M. Veltman, ``Diagrammar'', CERN-73-09.
\bibitem{42} O. Piguet \& S.P. Sorella, {\it Algebraic Renormalization}, Lect. 
Notes Phys. {\bf M28} (1995), 1.
\bibitem{43} A.D. Kennedy, J. Math. Phys. {\bf 22} (1981), 1330.
\bibitem{44} A. Bondi, G. Curci, G. Paffuti \& P. Rossi, Ann. Phys. {\bf 199}
(1990), 268.
\bibitem{45} A.N. Vasil'ev, S.\'{E}. Derkachov \& N.A. Kivel, Theor. Math.
Phys. {\bf 103} (1995), 487.
\bibitem{46} A.N. Vasil'ev, M.I. Vyazovskii, S.\'{E}. Derkachov \& N.A. Kivel,
Theor. Math. Phys. {\bf 107} (1996), 441.
\bibitem{47} A.N. Vasil'ev, M.I. Vyazovskii, S.\'{E}. Derkachov \& N.A. Kivel,
Theor. Math. Phys. {\bf 107} (1996), 710.
\bibitem{48} S. Laporta, Int. J. Mod. Phys. {\bf A15} (2000), 5087.
\bibitem{49} C. Studerus, Comput. Phys. Commun. {\bf 181} (2010), 1293.
\bibitem{50} C.W. Bauer, A. Frink \& R. Kreckel, cs/0004015.
\bibitem{51} P. Nogueira, J. Comput. Phys. {\bf 105} (1993), 279. 
\bibitem{52} J.A.M. Vermaseren, math-ph/0010025.
\bibitem{53} M. Tentyukov \& J.A.M. Vermaseren, Comput. Phys. Commun. {\bf 181}
(2010), 1419.
\bibitem{54} S.G. Gorishny, S.A. Larin, L.R. Surguladze \& F.K. Tkachov,
Comput. Phys. Commun. {\bf 55} (1989), 381.
\bibitem{55} S.A. Larin, F.V. Tkachov \& J.A.M. Vermaseren, ``The Form version
of Mincer'', NIKHEF-H-91-18. 
\bibitem{56} B.S. DeWitt, Phys. Rev. {\bf 162} (1967), 1195.
\bibitem{57} G. 't Hooft, Acta Universitatis Wratislaviensis {\bf 368} (1976), 
345, Proceedings of the 1975 Winter School of Theoretical Physics held in 
Karpacz.
\bibitem{58} B.S. DeWitt, in Proceedings of Quantum Gravity II, eds C. Isham, 
R. Penrose \& S. Sciama, (Oxford, 1980), 449.  
\bibitem{59} D.G. Boulware, Phys. Rev. {\bf D23} (1981), 389.  
\bibitem{60} L.F. Abbott, Nucl. Phys. {\bf B185} (1981), 189.
\bibitem{61} D.M. Capper \& A. MacLean, Nucl. Phys. {\bf B203} (1982), 413.
\bibitem{62} A.G.M. Pickering, J.A. Gracey \& D.R.T. Jones, Phys. Lett. 
{\bf B510} (2001), 347. 
\bibitem{63} O. Piguet \& S.P. Sorella, {\it Algebraic Renormalization}, Lect. 
Notes Phys. {\bf M28} (1995), 1.
\bibitem{64} J.A. Gracey, Phys. Lett. {\bf B525} (2002), 89. 
\bibitem{65} R.E. Browne \& J.A. Gracey, Phys. Lett. {\bf B540} (2002), 68. 
\end{thebibliography}
\end{document}